\documentclass[preprint,12pt]{elsarticle}

\usepackage{amssymb}
\usepackage{amsmath}

\usepackage{graphicx}
\usepackage{amsmath}
\usepackage[version=4]{mhchem}
\usepackage{siunitx}
\usepackage{longtable,tabularx}
\usepackage{subcaption}
\usepackage{hyperref}
\usepackage{textcomp}
\usepackage{multirow}
\usepackage{algorithm}
\usepackage{algpseudocode}

\algnewcommand\algorithmicforeach{\textbf{foreach}}
\algdef{S}[FOR]{ForEach}[1]{\algorithmicforeach\ #1\ \algorithmicdo}

\journal{Journal of Computational Science}

\begin{document}

\begin{frontmatter}



\title{Parallelizing Legacy Mesh Generation Software: Lessons Learned from a Pseudo-Constrained Parallel Data Refinement Approach for Advancing Front Local Reconnection}


\author[ODU]{Kevin Garner
\corref{corresponding-author}
} 
\author[MSU]{David Marcum} 
\author[ODU]{Nikos Chrisochoides
\corref{corresponding-author}
} 

\affiliation[ODU]{organization={Center for Real-time Computing, Old Dominion University},
            addressline={4700 Elkhorn Ave}, 
            city={Norfolk},
            postcode={23529}, 
            state={VA},
            country={USA}}
\affiliation[MSU]{organization={Center for Advanced Vehicular Systems, Mississippi State University},
            addressline={200 Research Blvd}, 
            city={Starkville},
            postcode={39759}, 
            state={MS},
            country={USA}}

\cortext[corresponding-author]{Corresponding authors. \textit{E-mail addresses:} kgarner0918@gmail.com (K. Garner) and nikos@cs.odu.edu (N. Chrisochoides).}

\begin{abstract}
This paper presents lessons learned from parallelizing the legacy software known as Advancing Front Local Reconnection (AFLR) as a black box. The parallel procedure utilizes (i) a data decomposition scheme where each subdomain is refined in parallel using the sequential mesh generation code and (ii) a runtime system for work-load balancing. Results on the mesh refinement operation show that the parallel method's stability (output mesh quality) is good and that the parallel method outperforms the serial AFLR by about 11 times when utilizing 16 CPU cores. However, full stability (i.e., generating the same quality as the serial method) and potential scalability is limited due to the constraints set by the black box's input boundary requirement. Satisfying this requirement for each subdomain not only adds overhead but also causes the parallel method to generate a different output mesh volume than that generated by the serial method. The complexity of such a state-of-the-art code requires that it be modified to a non-trivial extent in order to remove these constraints. These results suggest that the parallelization of black-box legacy codes like AFLR may not be practical and instead encourages an approach that is originally designed without such constraints to efficiently leverage the concurrency offered by large-scale architectures.
\end{abstract}




\end{frontmatter}


\section{Introduction} \label{aflr_introduction}
Mesh generation software is used in many industries where high-fidelity simulations are required, such as in healthcare, defense, and aerospace. For the last 30 years, many sequential Finite Element (FE) mesh generation methods and software were typically developed with a focus on functionality and optimized for single core architectures, without any thought towards scalability. NASA’s Computational Fluid Dynamics (CFD) 2030 Vision requires that highly functional mesh generation codes be capable of running on large-scale parallel architectures \cite{CFD2030}. Additionally, adaptive mesh techniques offer great potential, but have not seen widespread use due to issues related to software complexity, inadequate error estimation capabilities, and in some instances, complex geometries. These issues make the parallelization of existing sequential state-of-the-art mesh generation codes a highly preferable avenue for meeting the requirements of high-fidelity simulations.

Parallel mesh generation codes are typically developed using either one of the following approaches – functionality-first or scalability-first. The functionality-first approach attempts to parallelize existing state-of-the-art serial software that are fully functional (robust in its features and capabilities). The scalability-first approach focuses on designing the software from the ground up to maintain good scalability with the caveat of incomplete functionality. New features and capabilities are implemented on an as-needed basis. The functionality-first approach becomes viable if the code generates correct results (comparable to that generated by the serial code) and achieves a good speedup. Advancing Front Local Reconnection (AFLR) \cite{MarcumAFLR} is the serial code that is utilized in this parallelization effort and is a top, industrial-strength, sequential mesh generator that is used by NASA, the DoD, DoE, and several aerospace industry research groups. 
One must consider several attributes when designing a parallel mesh generation code that can offer the same features and capabilities of a sequential code while maintaining good parallel performance. The following attributes are critical not only for the successful execution of the parallel software but also for enhancing its adaptability to emerging high performance computing architectures, thereby ensuring its longevity and reliability when exploiting new methods for concurrency:

\begin{enumerate}
    \item \textbf{Stability} represents the level of quality of a mesh generated in parallel in comparison to the quality of a sequentially generated mesh. This quality is defined by the shape of individual elements and the number of elements within the mesh (fewer is better for the same shape constraint).
    \item \textbf{Robustness} is the ability of the parallel software to correctly and efficiently process the same input data as the equivalent sequential codes. The processing of data should not require operator intervention, as this is not only highly expensive for massively parallel computations, but most likely infeasible due to the large number of concurrently processed sub-problems. 
    \item \textbf{Scalability} is the ratio of the runtime of the best sequential implementation to the runtime of the parallel implementation, which should show a speedup that is ideally proportional to the number of processing elements utilized. However, this speedup is limited by the inverse of the sequential fraction of the software, according to Amdhal’s law \cite{AmdahlsLaw}. Therefore, non-trivial stages of the computation must be parallelized if one is to leverage as much concurrency as possible in the current and emerging architectures that are designed to deliver million- to billion-way concurrency. 
    \item \textbf{Code re-use} suggests a modular design that utilizes mesh operations (such as point creation/insertion, edge swapping, point smoothing, etc.) from any sequential (or previously-built parallel) mesh generators without requiring significant updates to accommodate these new operations. Rewriting new parallel algorithms for every new meshing operation can be highly expensive in time investment due to the complexity and evolving functionality of such codes. The code re-use approach is only feasible if the software satisfies the reproducibility criterion. 
    \item \textbf{Reproducibility} \cite{Chrisochoides18PDR} is a requirement that the mesh generator, when executed with the same input, produce either identical results (termed Strong Reproducibility) or those of the same quality (Weak Reproducibility) under the following modes of execution: (i) continuous without restarts, and (ii) with restarts and reconstructions of the internal data structures. This is essential when mesh elements undergo refinement more than once amongst different parallel processes, requiring the migration and reconstruction of their corresponding data structures when executing the mesh generation code.
\end{enumerate}

Although the parallel method shows good stability and outperforms the serial AFLR in terms of elements refined per second (across various core configurations), the method does not generate the same mesh quality as the serial method and its potential scalability is limited due to the constraints set by AFLR's input boundary requirement. Although the serial AFLR method satisfies the weak reproducibility criterion (shown in section \ref{background}), the parallel method fails the code re-use criterion due to the fact that its parallel framework underwent some modifications to adhere to AFLR's input requirement. Satisfying this requirement for each subdomain not only adds overhead but also causes the parallel method to generate a different output mesh volume than that generated by the serial method (even when utilizing some of AFLR's internal functionality to designate the same refinement settings for each subdomain as that in the serial approach, discussed in section \ref{analysis}). These results suggest that the parallelization of black-box legacy codes like AFLR may not be practical, as the complexity of such a state-of-the-art code requires that it be modified to a non-trivial extent in order to remove these constraints. We present several lessons learned from this effort and discuss needed future work for either the development of an unconstrained parallel AFLR method or a scalability-first approach.

The contributions of this paper include:

\begin{enumerate}
    \item the parallelization of the state-of-the-art sequential Advancing Front Local Reconnection software which maintains good output mesh quality and outperforms the serial method by about 11 times when utilizing 16 CPU cores and
    \item lessons learned regarding the utilization of such a highly functional sequential software as a black box within a parallel framework, including its feasibility in generating results comparable to the serial code and potential scalability when the method is left unmodified.
\end{enumerate}

Section \ref{background} provides an overview of the AFLR method and the parallel framework in which it is integrated. Section \ref{aflr_related_work} describes related works seen in literature. Section \ref{aflr_method} gives details of how the sequential method is integrated into the parallel framework. A performance evaluation and analysis is provided in section \ref{pscaflr_performance_evaluation}. Section \ref{future_work} discusses future work needed to develop either an unconstrained parallel AFLR or a scalability-first approach. Finally, section \ref{conclusion} concludes the paper.

\section{Background} \label{background}
The Center for Real-time Computing (CRTC) at Old Dominion University (ODU) previously proposed the telescopic approach \cite{Chrisochoides2016TelescopicAF}, a framework that is designed to exploit concurrency at multiple hardware levels for parallel grid generation. At the chip and node levels, the telescopic approach deploys a Parallel Optimistic (PO) \cite{DrakopoulosCDT3D, EwCCDT3D} layer and Parallel Data Refinement (PDR) layer \cite{ChernikovPDR2D2004, ChernikovDistributedDelaunay2006, ChernikovPDR3D2008}, respectively. The goal of this proposed effort was to (i)  parallelize AFLR at the PDR layer and if successful, (ii) to later transform AFLR using a speculative execution model for its integration at the PO layer. In contrast to the PO layer where concurrency varies, PDR maintains a fixed level of concurrency while parallelizing the refinement process. Its methodology is theoretically proven to maintain stability and robustness for parallel isotropic Delaunay-based mesh generation and has been experimentally verified for isotropic meshes \cite{ChernikovPDR2D2004, ChernikovDistributedDelaunay2006, ChernikovPDR3D2008}.

\begin{figure}[h!]
     \centering
     \begin{subfigure}[htb]{0.25\textwidth}
         \centering
         \includegraphics[width=\textwidth]{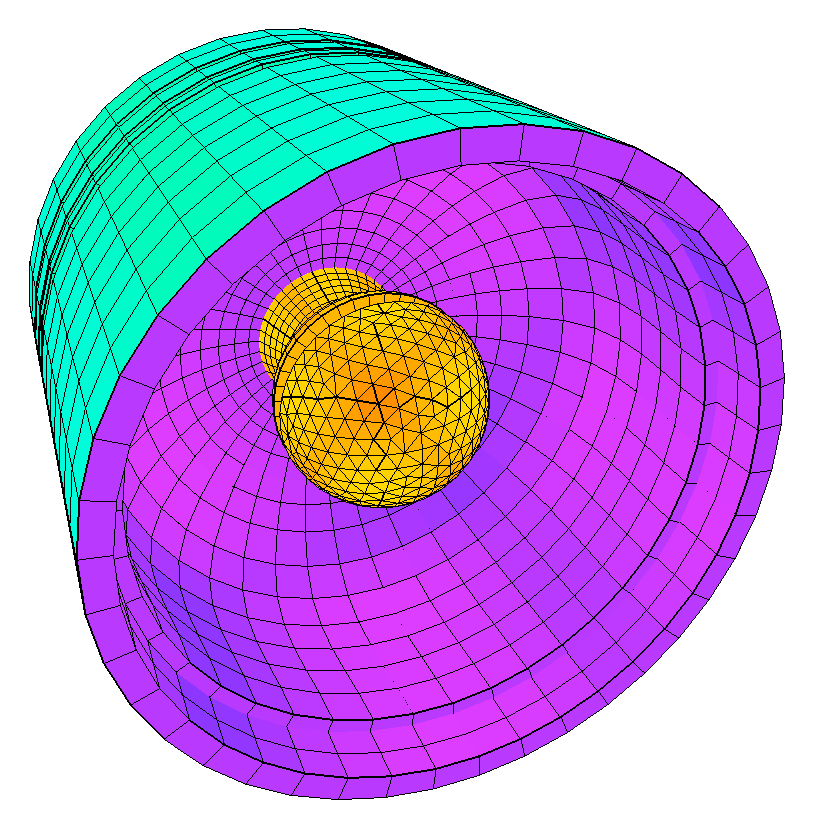}
         \caption{}
         \label{fig:plug_pic}
     \end{subfigure}
     \hfill
     \begin{subfigure}[htb]{0.7\textwidth}
         \centering
         \includegraphics[width=\textwidth]{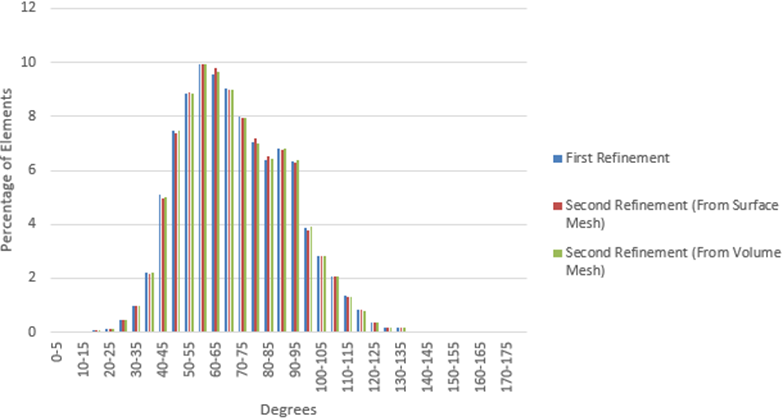}
         \caption{}
         \label{fig:plug_reproducibility}
     \end{subfigure}
        \caption[AFLR Plug Reproducibility]{AFLR reproducibility results for a plug geometry (a) are shown with regards to dihedral angle statistics (b) of the output meshes.}
        \label{fig:aflr_plug_reproducibility}
\end{figure}
\begin{figure}[h!]
     \centering
     \begin{subfigure}[htb]{0.2\textwidth}
         \centering
         \includegraphics[width=\textwidth]{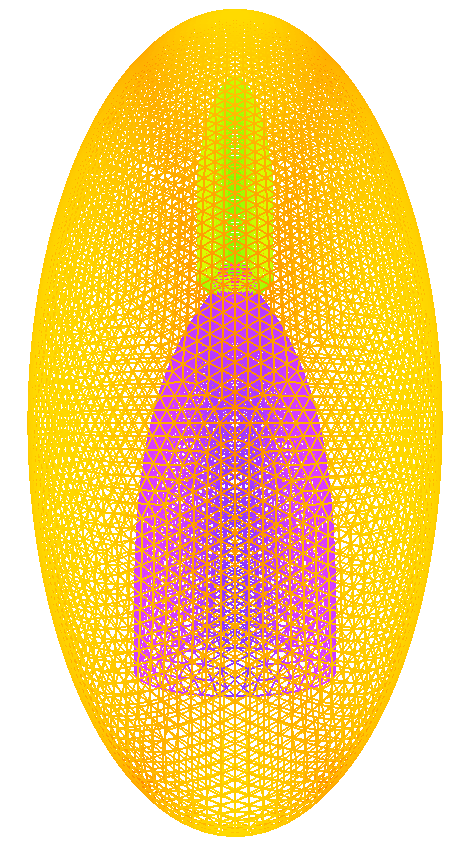}
         \caption{}
         \label{fig:missile_pic}
     \end{subfigure}
     \hfill
     \begin{subfigure}[htb]{0.75\textwidth}
         \centering
         \includegraphics[width=\textwidth]{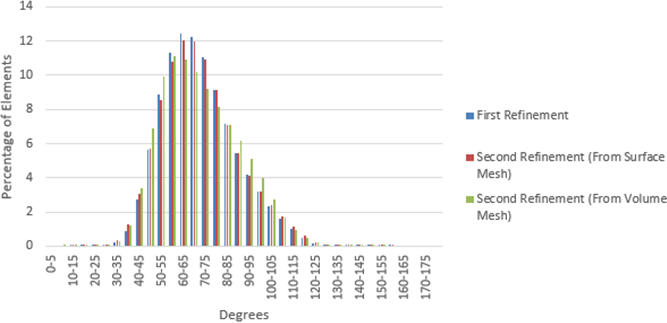}
         \caption{}
         \label{fig:missile_reproducibility}
     \end{subfigure}
        \caption[AFLR Missile Reproducibility]{AFLR reproducibility results for a missile geometry (a) are shown with regards to dihedral angle statistics (b) of the output meshes.}
        \label{fig:aflr_missile_reproducibility}
\end{figure}

Previous work involving the integration of the mesh generation software TetGen \cite{TetGen2015} with PDR shows that if the mesh generator fails to meet the reproducibility criterion in distributed memory, then the complexity of such state-of-the-art codes inhibits their modifications to a degree that their integration with parallel frameworks like PDR becomes impractical \cite{Chrisochoides18PDR}. The original, sequential AFLR software was determined to potentially be suitable for integration with PDR, as it was tested and shown to maintain weak reproducibility. Some results for a few selective geometries are shown in Figures \ref{fig:aflr_plug_reproducibility} and \ref{fig:aflr_missile_reproducibility}. Comparisons of the dihedral angle quality statistics are given by the output mesh (after the initial refinement) and the subsequent output meshes (one from the refinement of using the surface of the initial output mesh as input and another using the volume of the initial output mesh). While results show that the quality is comparable between all of the output meshes, it is not identical. Therefore, one can conclude that AFLR has weak reproducibility, which satisfies the reproducibility requirement for the PDR layer of the telescopic approach.

\subsection{Advancing Front Local Reconnection (AFLR)} \label{aflr_background}
In the case of three-dimensional unstructured grid generation, the AFLR software only accepts an input geometry with an established boundary surface grid (e.g., a triangulation, quadrilation, etc.). A Delaunay-based method is used to construct an initial boundary-conforming tetrahedral mesh (for a surface triangulation, for example). Each initial boundary point is assigned a value, by a point distribution function, representative of the local point spacing on the boundary surface. This function is used to control the final field point spacing. After the initial boundary-conforming tetrahedral mesh is constructed, numerous grid generation passes (i.e., iterations) are executed where each iteration involves point creation/insertion and local reconnection operations. All elements are initially made active, meaning that they need to be refined. If the points of an element satisfy the point distribution function, the element is made inactive and does not need to be refined. The advancing front method is used on active elements. A face of the element that is adjacent to another active element is selected. A new point is created by advancing in a direction, normal to the selected face, a distance that would produce an equilateral element based on an appropriate length scale (using the average point distribution). If a new point is too close to an existing point or another new point, it is rejected and removed. Accepted points are inserted into the existing grid by subdividing their containing elements. For example, if an edge point is inserted, then all elements sharing that edge are split. If a face point is inserted, then both elements sharing that face are split into three elements. All elements modified by point insertion, or any that undergo reconnection, are classified as active. A local reconnection operation is used to optimize the connections between points (or edges). Edges are repeatedly reconnected, or swapped, to satisfy a desired quality criterion. A min-max (minimize the maximum angle) criterion is primarily used which maximizes the minimum element edge weight, thereby producing high overall grid quality and eliminating most field sliver elements. After the grid generation passes have completed, all active elements undergo a final optimization phase, which consists of three quality improvement operations (sliver removal, grid coordinate smoothing, and further reconnection) \cite{MarcumAFLR}.

\subsection{Parallel Data Refinement (PDR)}
\begin{figure}[h!]
\begin{center}
\includegraphics[width=0.3\textwidth]{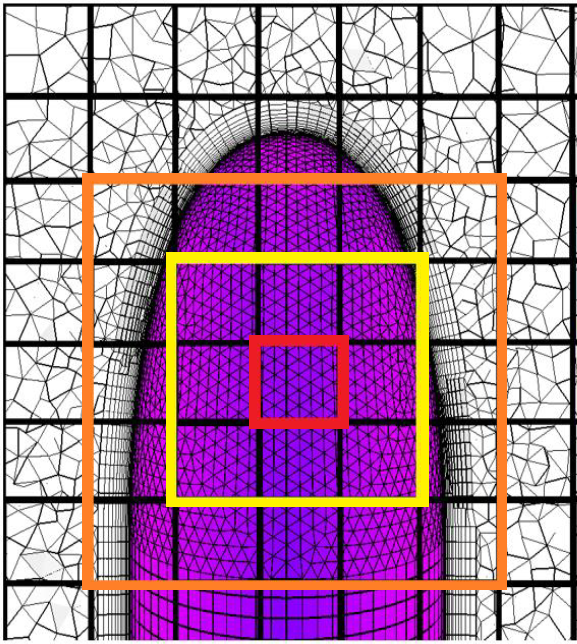}
\end{center}
\caption[2D PDR Example]{Shown is the upper portion of a PDR data-decomposed 3D rocket mesh (projected onto a 2D plane) where the red-boxed leaf is the primary leaf under refinement. The level 1 neighbors are those inside the yellow box (excluding the red leaf) and the level 2 neighbors are those inside the orange box (excluding all leaves inside the yellow box).}
\label{fig:PDR_Data_Decomposition}
\end{figure}

PDR decomposes a volume mesh by using an octree consisting of numerous leaves, or subdomains, that each contain a portion of the mesh. The general idea of PDR is to concurrently refine the octree leaves while maintaining mesh conformity. The main concern when parallelizing the refinement of a volume mesh are the creation of data dependencies caused by concurrent point insertions (when a thread attempts to insert a point within the leaf of another thread, for example, while the other thread is doing the same simultaneously), which can introduce mesh nonconformity between leaves if not handled through some synchronization method. PDR addresses this issue by introducing a buffer zone around each octree leaf. If a part of the mesh associated with a leaf is scheduled for refinement by a thread, no other thread can refine the parts of the mesh associated with the buffer zone of this leaf. This eliminates any data dependency risks and allows PDR to avoid fine-grain synchronization overheads associated with concurrent point insertions. Figure \ref{fig:PDR_Data_Decomposition} depicts an example (with a three-dimensional mesh projected onto a two-dimensional plane) of a leaf refinement by a single AFLR thread using PDR’s data decomposition, where a primary leaf under refinement and its corresponding buffer zone are identified. For a mesh M, we denote the set of all tetrahedra \(T\) within a leaf \(L_i\) and its neighboring regions \(N_1\) and \(N_2\) as follows:

\begin{equation}
\begin{split}
& T(L_i) = {\forall t \in M : b(t) \in L_i} \\
& N_1(L_i) = {\forall L : L \neq L_i \wedge \partial L \cap \partial L_i \neq \emptyset} \\
& T(N_1(L_i)) = {\forall t \in M : b(t) \in N_1(L_i)} \\
& N_2(L_i) = {\forall L : L \notin \{{L_i} \cup N_1(L_i)\} \wedge (\exists L_t \in N_1(L_i) : L \in N_1(L_t))} \\
& T(N_2(L_i)) = {\forall t \in M : b(t) \in N_2(L_i)}
\end{split}
\end{equation}

where \(N_1(L_i)\) and \(N_2(L_i)\) are the sets of level one and level two neighbors of leaf i, respectively, b(t) is the barycenter of tetrahedron t, and \(\partial L_i\) is the boundary of leaf i. We define a boundary/interface surface to be only the outermost faces of its corresponding region (termed an internal interface surface if surrounding a primary leaf or its level one neighbors, and an outer interface surface surrounding level two neighbors). If the level one neighbors and the primary leaf ($L_i$) are examined together in isolation, for example, the level one neighbor internal interface surface would contain every face that is connected to at most one tetrahedron within the set \(\{T(N_1(L_i)) \cup T(L_i)\}\). The level one neighbors of a leaf are considered to be the buffer zone of that leaf (no leaf in the buffer zone is scheduled for refinement while the primary leaf undergoes refinement).

\subsection{Pseudo-Constrained PDR}
When decomposing a volumetric mesh, subdomain interface surfaces are extracted (from some initial coarse mesh that was generated). The volume within each subdomain must conform to their respective interface surfaces. This not only adds overhead to the runtime of the parallel software, but constrains the overall meshing method  due to the fact that: (1) the boundary node distribution is often used to control volume point distribution (which affects the overall quality and density of elements within the mesh) and (2) these boundaries are used to maintain conformity between subdomains. This is a widely known problem that many research groups have addressed \cite{ZagarisVGRID, Lohner2014RecentAdvances, ParkRefine, EPIC2012, LOSEILLEFefloa}, where some solutions have been developed in order to mitigate the effects of these interfaces on the density and quality of the final mesh generated (detailed in section \ref{aflr_related_work}). However, a scalable solution has not yet been developed, to the best of the authors’ knowledge, that fully alleviates this problem.

PDR is designed to circumvent this issue by focusing solely on data decomposition while utilizing volume refinement methodologies \cite{Chrisochoides18PDR} and originally does not integrate octree leaf boundaries into the refinement process. However, due to the nature and design of AFLR, the PDR algorithm was modified in order to accommodate AFLR’s input requirement of an established surface grid. Consequently, its performance is constrained by the internal interface elements that are required to refine each subdomain. Some performance results and an in-depth analysis of the impact of these elements is given in section \ref{pscaflr_performance_evaluation}. Due to these required modifications and the subsequent constraints, this implementation diverted from the original PDR approach and will consequently be referred to as the Pseudo-Constrained Parallel Data Refinement for AFLR (or PsC.AFLR).

\section{Related Work} \label{aflr_related_work}
While there are several scalability-first methods in literature (such as \cite{NAVE2004191, ParkRefine, DrakopoulosCDT3D}), we specifically focus on the three-dimensional isotropic tetrahedral parallel mesh generation methods that follow the functionality-first approach (as does PsC.AFLR). One such effort integrated the binary of a sequential software, known as VGRID, into a distributed memory method. The method's performance suffered from several drawbacks, including insufficient exploitable concurrency due to the domain decomposition method utilized, VGRID's process creation, file I/O, and data structure initialization. As mentioned previously with regards to its parallelization effort, TetGen failed the reproducibility criterion (it could not successfully initialize its data structures based on a mesh that it itself had already generated) \cite{Chrisochoides18PDR}. The experiences in \cite{PDRPODMExperience} highlights several issues when integrating a Delaunay-based shared memory method known as PODM as a black box into the PDR framework. This PDR.PODM method would migrate large amounts of data (entire subdomains) throughout execution in order to resolve data dependencies that were required to satisfy the Delaunay property. Unfortunately, this overhead made the distributed approach 7x slower than the shared memory PODM when utilizing the same numbers of cores \cite{PDRPODMExperience}. The work in \cite{APrioriEdgesIso2015} involved the integration of the sequential software Netgen into a parallel framework called PMSH. While PMSH successfully generated multi-billion element meshes on large configurations of cores, the sequential software required modifications to successfully refine those geometries while some functionality was not successfully preserved in the parallel implementation. The approach in \cite{Pampa2017} provides a parallel framework called PAMPA which allows for the integration of a sequential mesher of the user's choice. While the user establishes qualitative criteria that PAMPA should follow for the refinement of the input mesh, subdomain boundaries must remain frozen during refinement until the entire mesh is re-partitioned for another refinement pass (this can occur over several iterations until most of the mesh elements satisfy qualitative criteria). The freezing of subdomain interface elements falls in line with AFLR's input boundary requirement; however, we predict that PAMPA would have similar issues (if using AFLR) as the presented PsC.AFLR approach when attempting to produce a desired mesh that adheres to user-provided quality criteria. AFLR's boundary requirement constrains the method even when manually specifying a point distribution for a subdomain's output volume (explored in section \ref{analysis}). Although their performance is hindered by global re-partitioning and mesh migration techniques, the methods in \cite{ParMMG2019} and \cite{MassivelyParallel2019} are capable of generating billion-element meshes. The method in \cite{ParMMG2019} takes a similar approach to \cite{Pampa2017} but instead uses the sequential MMG3D mesher to refine subdomains. 

The domain-defining grid (DDG) method in \cite{Lohner2014RecentAdvances} also uses an advancing front point placement technique to refine subdomains in parallel over numerous iterations. In contrast to PsC.AFLR, local reconnection is performed in the DDG method as a separate step after its iterations of point insertion have been completed. As specified in section \ref{aflr_background}, AFLR executes a combination of point insertion and reconnection operations within each of its grid generation iterations. Subdomain boundary elements are initially kept frozen in the DDG method and after each subdomain is refined in parallel, they are padded with extra layers of elements from neighboring subdomains and are then refined again. While this offers an improvement in the quality of these subdomain boundary elements and in turn, the quality of the final mesh, this process was reported to hinder the overall scalability of the algorithm due to the additional time required to move these layers and refine the interface elements for each subdomain \cite{LohnerIMR2013}. In order to control the point distribution within each subdomain, background grids, source points, and CAD-based information can all be used as input to the DDG method. A different approach is taken within PsC.AFLR, and the subsequent results (explored in section \ref{analysis}) demonstrate that if one cannot fully control subdomain point distribution, it can become an impediment in producing the desired mesh when using such an advancing front software as a black box within the parallel algorithm.

It should be noted that some of the aforementioned methods (such as \cite{Lohner2014RecentAdvances} and \cite{MassivelyParallel2019}) each built upon codes that have evolved with institutional memory; that is, their respective developers made the necessary modifications to these codes in order to accommodate their parallel algorithms and thus achieve the desired results for various use cases. This paper presents the lessons learned when attempting to parallelize a sequential black box mesh generation software, showing that if a black box code does not evolve to meet the required attributes of a parallel framework such as PDR, it may not achieve full stability and good scalability.

\section{Methodology} \label{aflr_method}
Two implementations were developed for PsC.AFLR, where the first attempts to improve the quality of each subdomain's internal interface elements throughout refinement and the second keeps these elements frozen throughout refinement until a final optimization pass is executed over the entire mesh. The latter method was shown to be the optimal approach due to its reduced overhead and the fact that it still maintains good output mesh quality compared to both the first approach and the serial AFLR software. A high-level algorithm of PsC.AFLR is shown in Algorithm \ref{alg:pscaflr_high_level_algorithm}. Both methods share the overall algorithm with one key difference - line 12, the execution of local reconnection over the union of the refined leaf and its level one neighbors (i.e., a super-subdomain). This reconnection serves as the avenue for improving the quality of internal interface elements in the first approach.


\begin{algorithm}
\caption{High-level Algorithm of PsC.AFLR (line 12 is included only in the super-subdomain reconnection implementation of the method)}\label{alg:pscaflr_high_level_algorithm}
\footnotesize{
\begin{flushleft}
PsC.AFLR(G, \textit{d}) \\
\textbf{Input}: G is the input geometry \\
\hspace*{\algorithmicindent} \ \ \ \ \ \textit{d} is the depth at which the PDR octree should be constructed \\
\textbf{Output}: Generated mesh that conforms to the input geometry's point distribution (subject to interface boundary constraints)
\end{flushleft}
\begin{algorithmic}[1]
\State Generate initial volume mesh for G
\State Construct octree around initial volume mesh with depth \textit{d}, creating n leaves
\State Assign all points and elements to octree leaves $L_1...L_n$
\State Insert leaves $L_1...L_n$ into refinement list R
\While{R is NOT empty}
\ForEach{leaf \textit{l} $\in$ R} \Comment{Lines 10-12 executed in parallel}
\If{level one neighbors $N_1(l)$ are NOT locked}
\State Remove \textit{l} from R
\State Lock $N_1(l)$
\State Extract internal interface surface S of \textit{l}
\State AFLR\_REFINE(S,\textit{l})
\State AFLR\_RECONNECTION(\textit{l} $\cup$ $N_1(l)$)
\State Unlock $N_1(l)$
\EndIf
\EndFor
\EndWhile
\State Merge $L_1...L_n$ into one mesh M
\State AFLR\_FINAL\_OPTIMIZATION(M)
\State TERMINATE()
\end{algorithmic}
}
\end{algorithm}

First, we describe in detail how the overall algorithm was implemented, as it pertains to both approaches. Then we describe the challenges and necessary measures taken to permit the execution of local reconnection over the interface elements. Corresponding performance results of each approach are given in section \ref{pscaflr_performance_evaluation}. In order to integrate AFLR into the pseudo-constrained PDR framework and utilize its functionality effectively, an intricate understanding of its underlying data structures and methodologies was required.

\begin{figure}[h!]
     \centering
     \begin{subfigure}[htb]{0.45\textwidth}
         \centering
         \includegraphics[width=\textwidth]{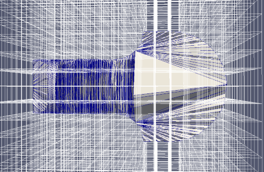}
         \caption{}
         \label{fig:initial_coarse_mesh_pic}
     \end{subfigure}
     \hfill
     \begin{subfigure}[htb]{0.49\textwidth}
         \centering
         \includegraphics[width=\textwidth]{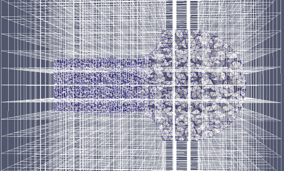}
         \caption{}
         \label{fig:initial_dense_mesh_pic}
     \end{subfigure}
        \caption[Coarse vs. Dense Mesh]{The difference is shown between an initial volume mesh generated for a horn bulb geometry that is too coarse (a) and one that is sufficiently dense (b) for an octree at depth level 4.}
        \label{fig:coarse_dense_mesh}
\end{figure}

The initial volume mesh is generated by TetGen \cite{TetGen2015} for PsC.AFLR. PsC.AFLR requires an initial volume mesh that is dense enough to satisfy boundary requirements of individual octree leaves (given that leaf boundaries are extracted from the faces of initial volume elements). AFLR requires a smooth, simply-connected surface when refining a domain (meaning that an input triangulation will not be accepted if it contains an edge that is shared by more than two faces). If the initial mesh is too coarse, there may be a tetrahedron that spans multiple leaves. In this scenario, no boundary can be extracted for each leaf if a face is spanning across them all. One solution would be to reduce the octree depth level (increasing the size of the leaves/subdomains), but this would reduce the overall number of leaves/subdomains, thereby reducing the amount of achievable concurrency. In order to maintain the desired level of concurrency, the initial mesh must be dense enough so that boundary faces can be extracted for each leaf. The difference between a mesh that is too coarse and one that is sufficiently dense for an octree at depth level 4 can be seen in Figure \ref{fig:coarse_dense_mesh}. A problem observed when using AFLR to generate an initial volume mesh is that it does not always generate tetrahedra within a certain volume constraint unless the mesh undergoes a significant amount of refinement. This is counterintuitive for the purpose of generating an initial mesh. If the initial mesh undergoes too much refinement (in order to satisfy the density requirement), then more runtime will be spent at this pre-processing stage rather than in parallel refinement, thereby rendering the parallel refinement futile. The Delaunay-based TetGen software is used to generate an initial volume mesh due to its ability to refine a geometry in short time using a strict volume constraint. 

\begin{figure}[h!]
\begin{center}
\includegraphics[width=0.6\textwidth]{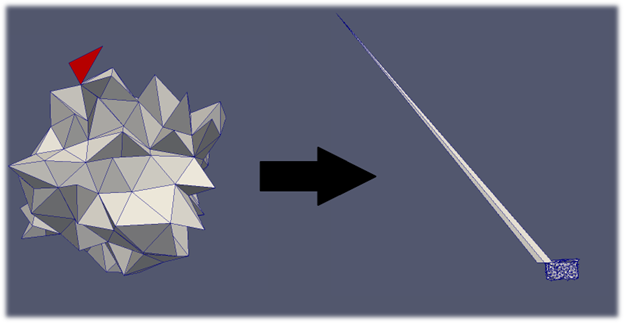}
\end{center}
\caption[Loosely-connected Partition]{The left side shows a subdomain with one tetrahedron (in red) connected to the other tetrahedra by only a point (loosely-connected). The right side shows a result generated after refining this subdomain (including the point-connected tetrahedron) with AFLR.}
\label{fig:loosely_connected_partition}
\end{figure}

After an octree is constructed for the TetGen-generated mesh, data are assigned to octree leaves based on element barycenter. If the barycenter for an element falls within a leaf, that element is assigned to that leaf. Once data assignment is complete, it is possible that some leaves will have loosely-connected partitions (such as in Figure \ref{fig:loosely_connected_partition}), that is, elements (or groups of elements) connected to another element (or group) by only a point or edge. This is not acceptable for AFLR. Every tetrahedron in the domain must be face-connected to another; otherwise, incorrect results can potentially be generated (such as that on the right in Figure \ref{fig:loosely_connected_partition}). The face connectivity of tetrahedra within a subdomain can be considered as an undirected graph, where a tetrahedron is a vertex and an edge denotes that two tetrahedra share a face. Through edge traversal, if every vertex of the undirected graph can reach every other vertex, then the subdomain has no loosely-connected partitions. To mitigate this issue, a check occurs before the refinement of a subdomain. If any loosely-connected tetrahedron is found, neighboring leaves are examined to find a tetrahedron with a matching face to the loosely-connected tetrahedron. Once found, the tetrahedron in question is reassigned to that particular leaf. This routine is repeated until all connectivity issues are resolved.

In PsC.AFLR, a boundary must be extracted for a subdomain before any refinement of its elements can begin. This introduces overhead to the overall process due to the requirement that a smooth, simply-connected surface must be given as input to AFLR. The set of tetrahedra within a leaf is examined in isolation (as if the subdomain is the entire domain). Any face that is not shared between tetrahedra is considered to be a boundary face. It is possible to extract a boundary that contains an edge which is shared by more than two faces. This is not acceptable for AFLR. This scenario occurs when there is a tetrahedron that has a barycenter just outside the octree leaf's dimensions, causing it to be assigned to a neighboring leaf, while its neighboring tetrahedra were assigned to the primary leaf. When such an edge is found in the extracted boundary, the corresponding tetrahedron is located within the neighboring leaf, removed from that leaf, and added to the primary leaf. The boundary is extracted and examined again. This routine repeats until a manifold boundary, acceptable for AFLR, is extracted.

After all leaves have undergone refinement, their data are merged into a single mesh for a final phase of quality improvement/optimization passes (similar to the serial AFLR's final optimization phase). A function was added to AFLR's API to perform only quality improvement/optimization (sliver removal, smoothing, and local reconnection) sequentially on a domain while skipping its mesh generation phase.

\subsection{Refinement of Internal Interface Elements} \label{method_internal_interface_refinement}
As mentioned previously, additional measures were taken to allow the refinement of subdomains' internal interface elements. If a boundary face (shared by two leaves) undergoes refinement, then the corresponding elements within both leaves must be updated (which adds dependencies and increases overall runtime due to the required communication between the corresponding threads). Otherwise, the connectivity between the subdomains will be incorrect and the final mesh will be non-conforming. Subdomain boundary refinement is considered so that boundary elements do not retain poor quality by the end of refinement. In implementing a solution, AFLR was modified to not only accept a single set of data (points, triangles, and tetrahedra) for one leaf, but to also accept a second set of data – the set of all of its level one neighboring leaves. During refinement of the primary leaf, the internal interface surface is kept frozen, meaning that point insertion is not allowed on the leaf's boundary. AFLR refines the individual leaf (advancing front point placement/insertion and local reconnection) but does not make any optimizations/quality improvement as the serial AFLR would. Instead, the newly refined leaf is merged with its level one neighbors into a super-subdomain and local reconnection is performed over the super-subdomain (thereby allowing the optimization of the primary leaf’s boundary elements). The internal interface between level one and level two neighbors remains frozen, so as to eliminate the need of updating level two neighbors during refinement (maintaining PDR’s original method of concurrency). 

It is possible to have duplicate points when merging the two sets of data (leaf and its level one neighbors) given that they share interface faces and conform to one another. When these sets of data are merged, duplicate points are removed and any tetrahedron or triangle that references these points are updated to use the same point identifiers. All tetrahedra and faces use integer-based indices in AFLR to denote which points they contain; for example, if two tetrahedra contain the same point, they will use the same integer-based index to reference that point. The removal of duplicate points is necessary as they are not permitted by AFLR. 

Once local reconnection over the super-subdomain has completed, this refined data is assigned to octree leaves. No points are deleted during refinement, so only new points are added to leaves. All new tetrahedra are assigned to leaves. Having undergone swapping, a tetrahedron will have a different barycenter. It is possible for the barycenter to move just enough to be assigned to a level two neighbor. If a level two neighbor must be updated during refinement, then this limits potential speedup and conflicts with PDR’s method of concurrency. A thread should only refine a leaf and its level one neighbors without allowing any changes to propagate beyond the level one region. If this situation occurs, the tetrahedron is assigned to the leaf that (1) contains a tetrahedron with a matching face and (2) is a level one neighbor of both the level two leaf and of the primary leaf.

An additional overhead involves the temporary removal of tetrahedra from a level one neighbor subdomain that contains level two points, and the later re-assignment of these tetrahedra back to their respective leaves after refinement. This was necessary as level one neighbor boundary elements may contain points that are assigned to level two neighbors, in which case these points were not packed and migrated to the high-performance computing node where the refinement process was scheduled to commence. Only a leaf and its level one neighbor data (that is, data within the level one neighbor leaves) are packed for migration, which follows PDR’s method of concurrency. The temporary removal of these tetrahedra are acceptable because these elements will eventually undergo local reconnection when their respective leaf is refined. 

\subsection{Integration of PsC.AFLR onto PREMA}
Parallelism was achieved by fully integrating PsC.AFLR onto a runtime system called the Parallel Runtime Environment for Multi-computer Applications (PREMA) 2.0 \cite{BarkerPREMA, Thomadakis18PREMA}. PREMA 2.0 is a parallel runtime system developed to support adaptive and irregular applications. It is capable of running in both shared and distributed memory. PREMA provides a globally addressable namespace, message forwarding, and data migration capabilities by using constructs called mobile objects and mobile pointers. Mobile objects are user-defined data objects that may encapsulate data not necessarily residing in contiguous memory (such as a leaf and its level one neighbors). A mobile pointer is a unique identifier created for each mobile object that can be used by the system even if the object has migrated to a different rank. This enables a rank to send a message to a specific mobile object and execute a user-defined function with that object, regardless of its location. In PsC.AFLR, a master-worker model is used, where lines 1-9 and 13-19 of the high-level algorithm (Algorithm \ref{alg:pscaflr_high_level_algorithm}) are executed by the master thread and lines 10-12 are executed by worker threads in parallel. During run time, the PsC.AFLR method exposes data decomposition information (number of leaves/subdomains waiting to be refined in the list) to the underlying run-time system. Pseudo-constrained PDR essentially informs PREMA that a leaf may undergo refinement if it and its level one neighbors are not currently under use in another leaf’s refinement process (i.e., are unlocked). The master thread will send a message to the corresponding mobile pointer (representative of the leaf and its set of neighbors that are ready for refinement), essentially informing PREMA to execute a refinement function given the mobile object’s data. PREMA monitors the load of the system and performs migration (of the leaf and neighbor data) to an available worker without interrupting execution. Communication and execution are separated into different threads to provide asynchronous message reception and instant computation execution at the arrival of new work requests.

\section{Performance Evaluation} \label{pscaflr_performance_evaluation}
PsC.AFLR was executed and tested on the Turing high-performance computing cluster at Old Dominion University \cite{ODUTuring}. Turing contains dual socket nodes that each feature two Intel Xeon E5-2698 v3 CPUs @ 2.30GHz with 128 GB of memory (32 cores total per node). GNU GCC version 11.4.1 and Intel MPI compilers were used for compilation and execution. All quantitative data presented are from the refinement time of both applications (PsC.AFLR and serial AFLR) and does not include initial volume mesh generation time or end optimization time. The qualitative data presented represent the final output meshes from end-to-end execution. Our goals were focused first on achieving full stability of PsC.AFLR and scalabilty of the refinement operation, then addressing its pre- and post-processing operations in future work. 

\subsection{Results}
\begin{figure}[h!]
\begin{center}
\includegraphics[width=0.65\textwidth]{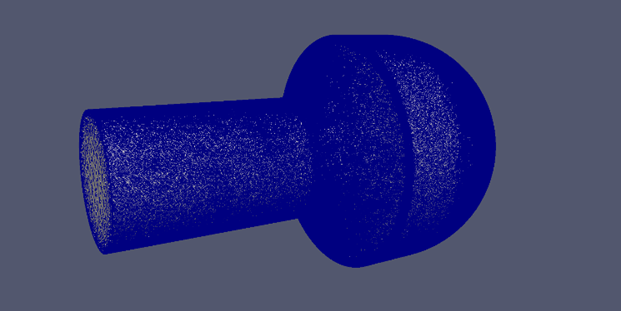}
\end{center}
\caption[Horn bulb Geometry]{Shown is a horn bulb geometry with 1,062,042 surface elements.}
\label{fig:horn_bulb_pic}
\end{figure}

Figure \ref{fig:horn_bulb_pic} shows a geometry of a horn bulb that was used for testing with both PsC.AFLR and serial AFLR. The horn bulb geometry contains 1,062,042 surface elements. The number of tetrahedra within the initial volume mesh generated by TetGen was 3,773,233 for PsC.AFLR. The octree used in PsC.AFLR is at depth level 4 (containing 4,096 leaves, or subdomains). 

\begin{figure}[h!]
\begin{center}
\includegraphics[width=1\textwidth]{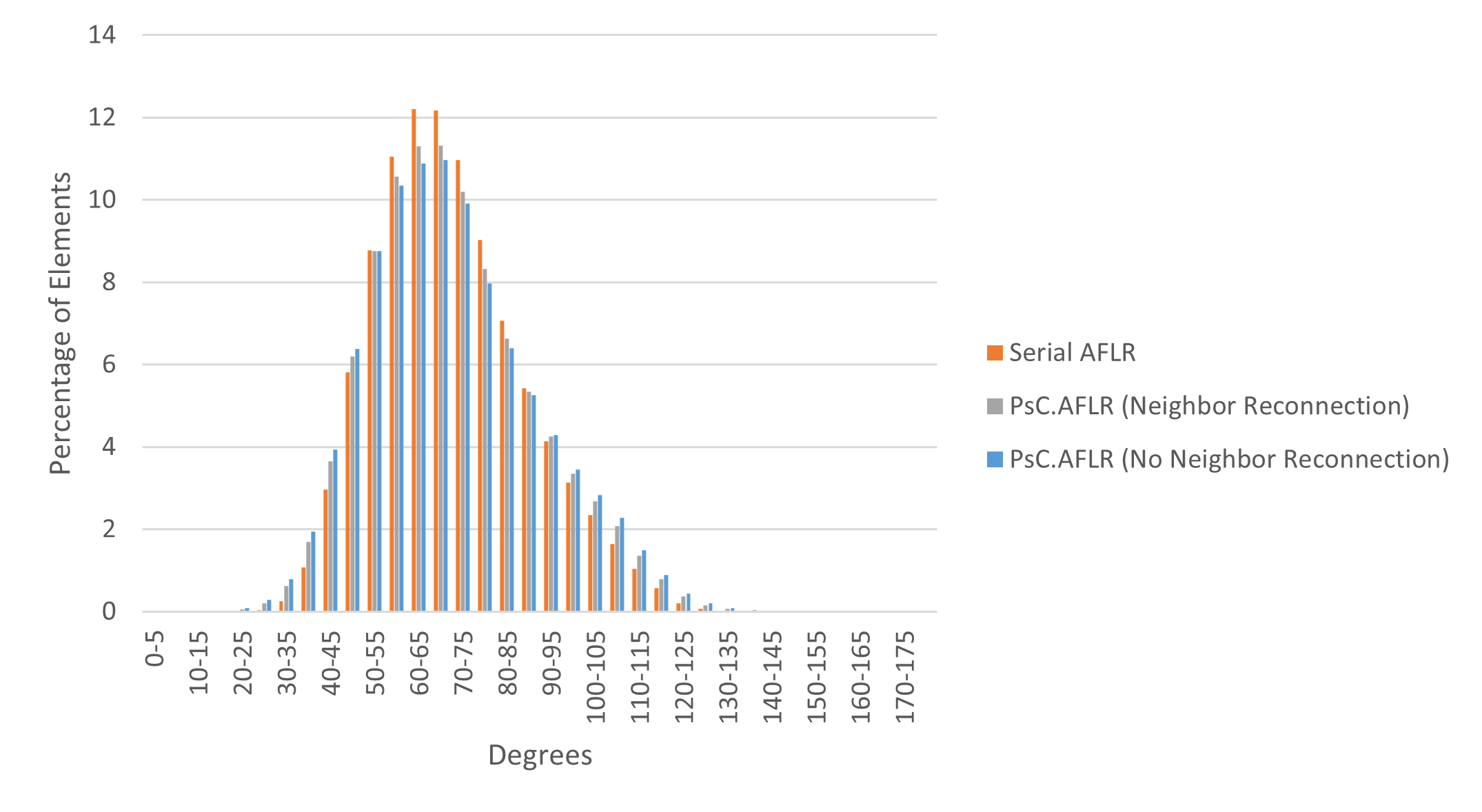}
\end{center}
\caption[Overall Dihedral Angle Statistics for Super-subdomain Reconnection]{The overall dihedral angle statistics with regards to performing super-subdomain local reconnection are shown.}
\label{fig:Horn_bulb_Neighbor_Reconnection_Overall}
\end{figure}

\begin{figure}[h!]
\begin{center}
\includegraphics[width=0.9\textwidth]{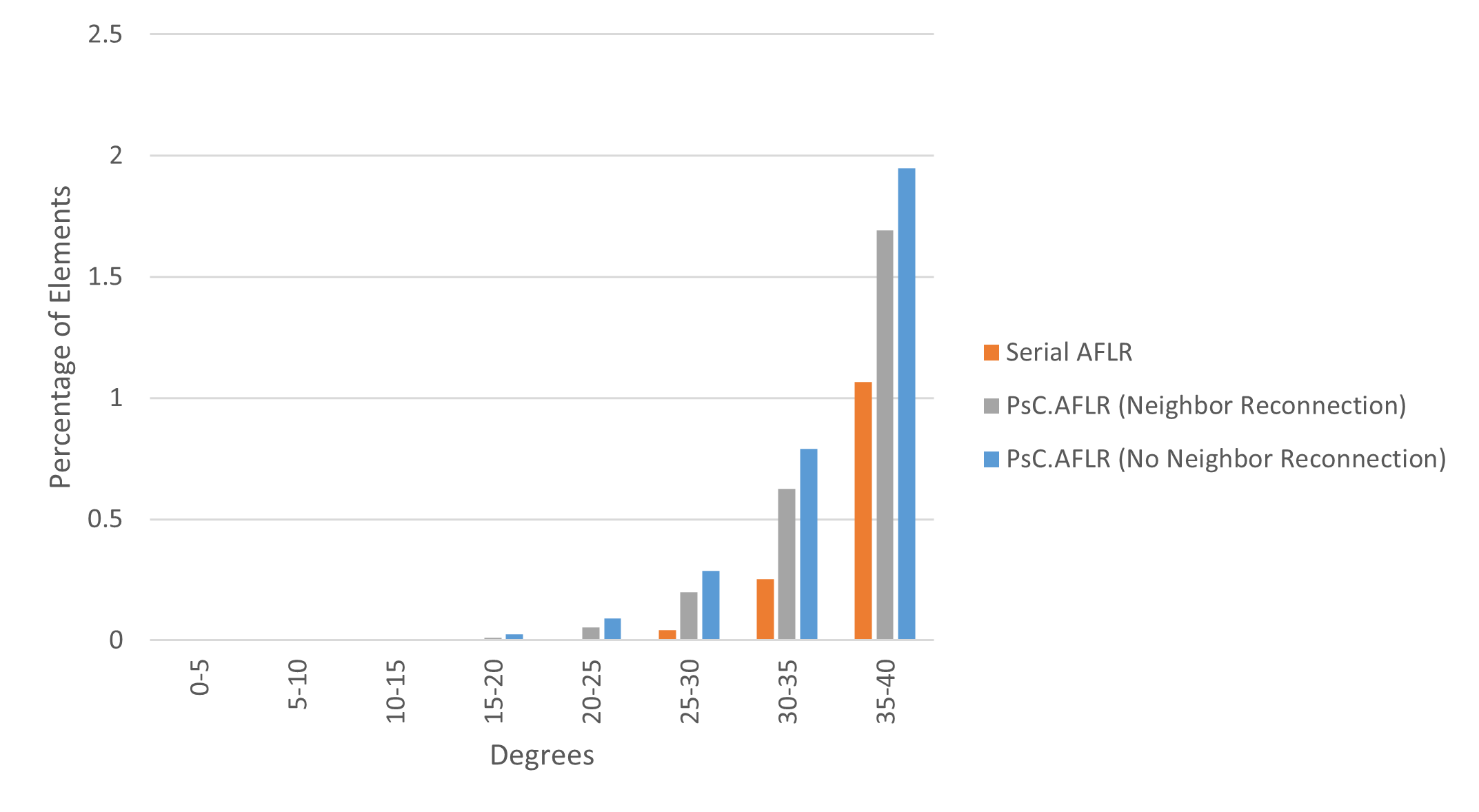}
\end{center}
\caption[Small Dihedral Angle Statistics for Super-subdomain Reconnection]{The small dihedral angle statistics with regards to performing super-subdomain local reconnection are shown.}
\label{fig:Horn_bulb_Neighbor_Reconnection_Small}
\end{figure}

\begin{figure}[h!]
\begin{center}
\includegraphics[width=0.9\textwidth]{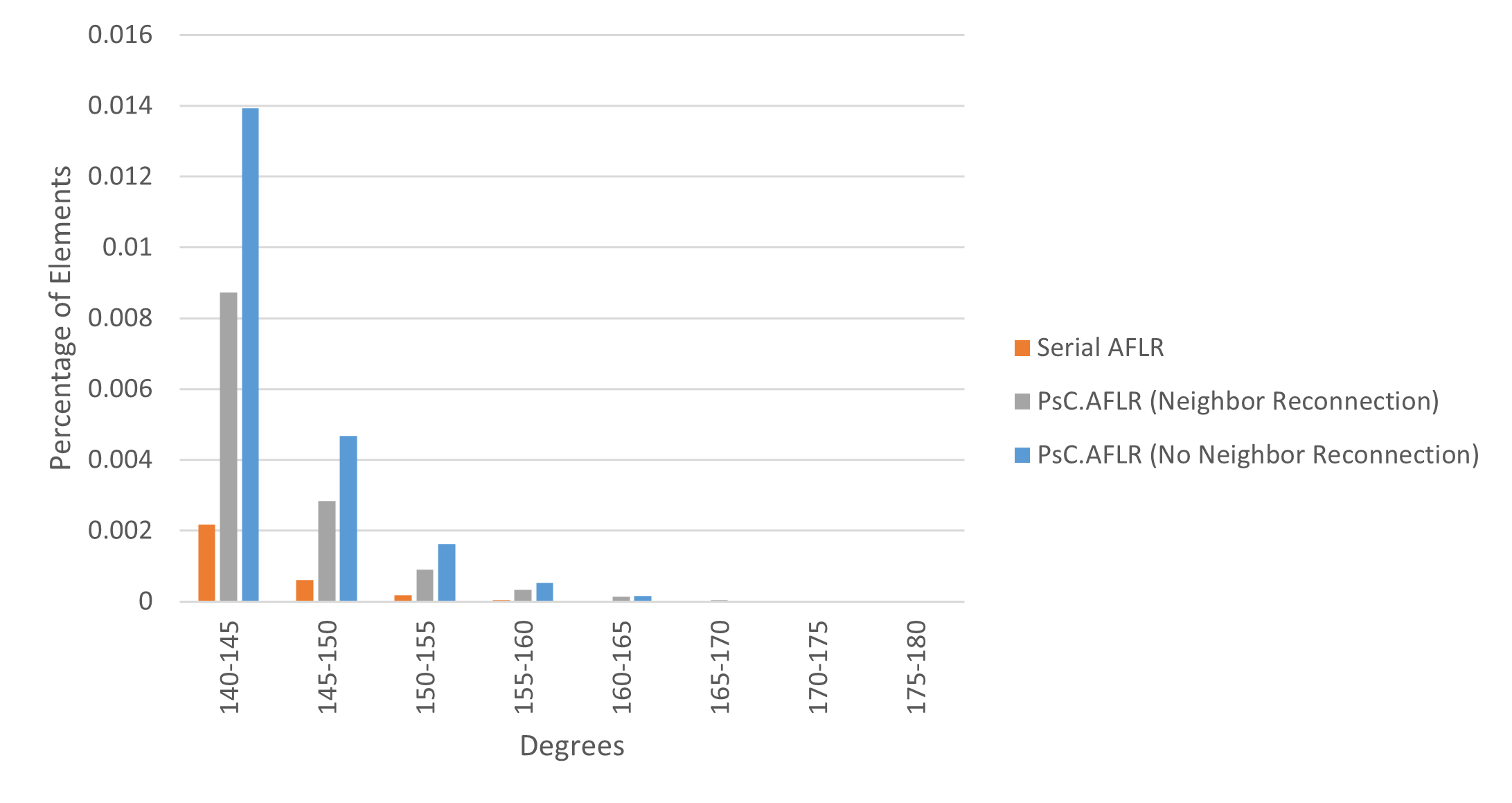}
\end{center}
\caption[Large Dihedral Angle Statistics for Super-subdomain Reconnection]{The large dihedral angle statistics with regards to performing super-subdomain local reconnection are shown.}
\label{fig:Horn_bulb_Neighbor_Reconnection_Large}
\end{figure}

As specified previously, one of PsC.AFLR's implementations involved merging level one neighbor leaves with their corresponding primary leaf after the primary leaf's refinement, then performing local reconnection over the super-subdomain to improve the quality of the primary leaf’s subdomain boundary elements. However, if internal interface elements remain frozen throughout refinement and local reconnection is not performed on these elements, the quality of the final mesh generated by PsC.AFLR is still comparable to that generated by the serial AFLR and falls within the operational limits of CFD solvers (such as FUN3D \cite{Fun3D} and SU2 \cite{SU2}). Figures \ref{fig:Horn_bulb_Neighbor_Reconnection_Overall}, \ref{fig:Horn_bulb_Neighbor_Reconnection_Small}, and \ref{fig:Horn_bulb_Neighbor_Reconnection_Large} compare the dihedral angle statistics of the meshes output when performing local reconnection over super-subdomains and without performing this operation (skipping line 12 in Algorithm \ref{alg:pscaflr_high_level_algorithm}). The geometry under refinement in this test is the horn bulb. The minimum dihedral angles of the meshes generated are 7.3, 3.47, and 3.52 degrees for serial AFLR, PsC.AFLR (with super-subdomain local reconnection), and PsC.AFLR (without super-subdomain local reconnection), respectively. The maximum dihedral angles are 164.57, 172.58, and 165.71 degrees for serial AFLR, PsC.AFLR (with super-subdomain local reconnection), and PsC.AFLR (without super-subdomain local reconnection), respectively. These results encouraged an implementation that excludes super-subdomain reconnection, and consequently a reduction in potential overhead that would be caused by the additional routines needed for super-subdomain reconnection (discussed in section \ref{method_internal_interface_refinement}). Because super-subdomain reconnection does not significantly improve output mesh quality or help with the problem of the parallel method generating a different volume (with different quality) than the serial method (discussed in section \ref{analysis}), we focus solely on the latter implementation (since it has good overall output quality that still falls within operational limits of CFD solvers). The inferior performance data related to the super-subdomain reconnection implementation can be found in \cite{GarnerThesis}.  

Table \ref{table:second_perf_table} shows the performance achieved by PsC.AFLR (with no super-subdomain reconnection routines), in addition to its refinement speed for each configuration of cores used. The final number of tetrahedra generated by PsC.AFLR is approximately 14 million and about 116 million are generated by serial AFLR. Serial AFLR’s refinement time is 16,101 seconds and its refinement speed is 7,212 elements/sec. PsC.AFLR is capable of generating a larger number of elements per second than the serial code on each configuration of cores (even on one core, which is explained in the analysis in section \ref{analysis}) and outperforms serial AFLR by about 11 times when using 16 CPU cores. Although PsC.AFLR produces a final mesh with fewer elements than the serial software, its final mesh maintains satisfactory quality in comparison, as shown previously in Figures \ref{fig:Horn_bulb_Neighbor_Reconnection_Overall}, \ref{fig:Horn_bulb_Neighbor_Reconnection_Small}, and \ref{fig:Horn_bulb_Neighbor_Reconnection_Large}. 

\begin{table}[h!]\scriptsize
\centering
\begin{tabular}{ccccccccc}
 \hline
 \# of Cores&\multicolumn{2}{c}{1}&2&4&8&16&32&64\\
 \hline
 \# of generated elements & \textcolor{red}{116M} & 14.6M & 14.6M & 14.4M & 14.7M & 14.4M & 14.4M & 14.5M \\
 Runtime (sec) & \textcolor{red}{16.1K} & 856 & 573 & 401 & 224 & 175 & 175 & 278\\
 Refine Speed (Elems/sec) & \textcolor{red}{7.2K} & 17.1K & 25.6K & 36K & 65.9K & 82.4K & 82.5K & 52.4K \\
 \hline
\end{tabular}%
\caption[PsC.AFLR Horn Bulb Performance Results]{Approximate PsC.AFLR horn bulb performance results are shown. The serial AFLR data is denoted in red while the remainder is of PsC.AFLR. `K' means thousand and `M' means million.}
\label{table:second_perf_table}
\end{table}



\begin{figure}[h!]
\begin{center}
\includegraphics[width=0.75\textwidth]{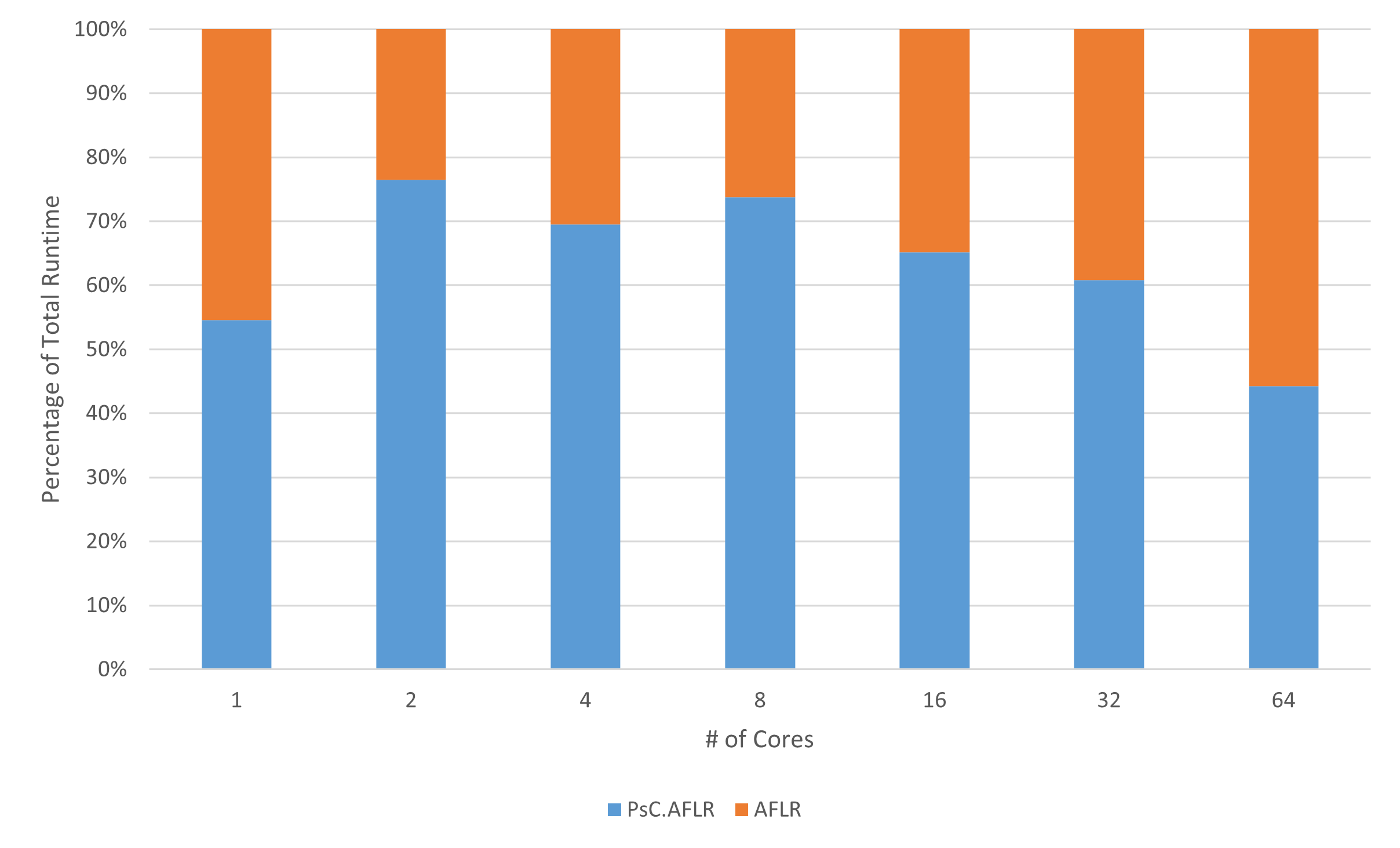}
\end{center}
\caption[PsC PDR vs. AFLR Profile]{A profile is shown of PsC PDR operations vs. AFLR operations for the horn bulb geometry.}
\label{fig:PsC_AFLR_Profile_Implementation_2}
\end{figure}
\begin{figure}[h!]
\begin{center}
\includegraphics[width=0.75\textwidth]{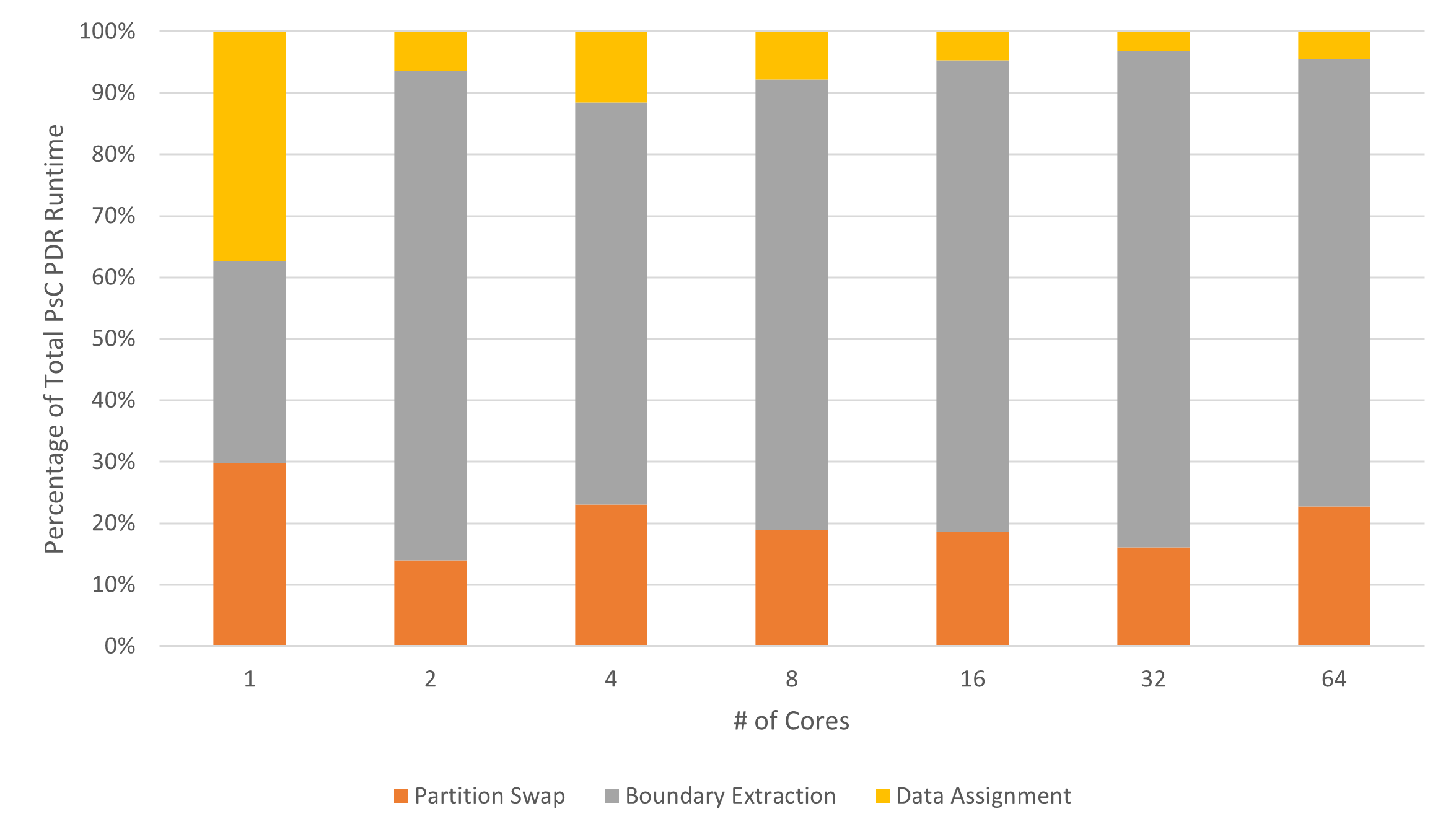}
\end{center}
\caption[PsC PDR Profile]{A profile is shown of PsC PDR operations for the horn bulb geometry.}
\label{fig:PsC_PDR_Profile_Implementation_2}
\end{figure}

A percentage breakdown of the average time spent executing PsC PDR operations and AFLR operations (within the context of PsC.AFLR) is shown in Figure \ref{fig:PsC_AFLR_Profile_Implementation_2}. The average time for each PsC operation was gathered, compared to the total time spent executing PsC-specific operations, in Figure \ref{fig:PsC_PDR_Profile_Implementation_2}. This data was gathered by taking the average times among the master and worker processes in 100 runs, executed for each configuration of cores (100 runs for 1 core, 100 runs for 2 cores, etc.). Figure \ref{fig:PsC_AFLR_Profile_Implementation_2} shows that the time spent executing PsC PDR operations varies on different configurations of cores but lessens as the numbers of cores are increased. Overall, PsC PDR operations dominate a percentage of the runtime. In Figure \ref{fig:PsC_PDR_Profile_Implementation_2}, Boundary Extraction (repeatedly checking that no boundary edge is shared by more than two faces and re-assigning elements between subdomains to address this issue) is the most time-consuming PsC PDR operation. Partition Swap, the second most expensive operation, represents the time spent detecting loosely-connected partitions and re-assigning them to neighboring leaves. Both of these expensive operations underline the fact that the decomposition technique utilized by PsC.AFLR can significantly impact potential overhead.

\begin{figure}[h!]
\begin{center}
\includegraphics[width=0.75\textwidth]{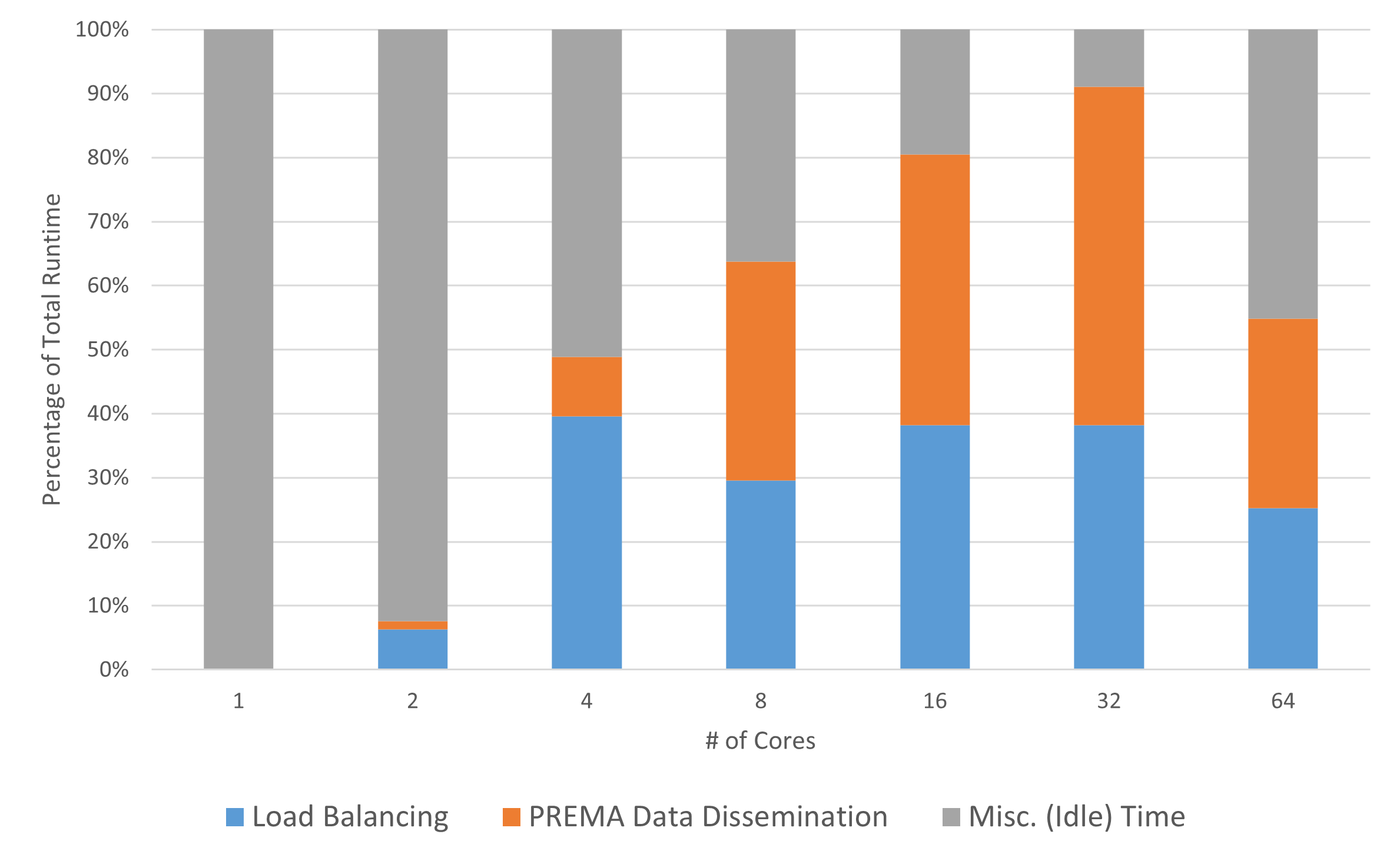}
\end{center}
\caption[PREMA \& Load Balancing Profile]{A profile is shown of PREMA \& load balancing routines for the horn bulb geometry.}
\label{fig:PREMA_Profile_Implementation_2}
\end{figure}

A breakdown of the time spent making data dissemination decisions by PREMA and time spent executing load balancing operations is shown in Figure \ref{fig:PREMA_Profile_Implementation_2}. Load balancing operations include the packing, unpacking, and migration of data between parallel processes. Data dissemination decisions made by PREMA account for the sending of messages between the master and worker processes, requests for work made by the workers, and the master determining which workers to assign work based on their current workloads. Between load balancing and PREMA, more time is spent in load balancing when PsC.AFLR is executed on a smaller number of cores while more time is spent making data dissemination decisions when the program is executed on a larger number of cores. This happens simultaneously while PsC PDR and AFLR operations are being executed, due to PREMA’s asynchronous message reception and computation execution being handled in separate threads. Although load balancing and PREMA do not make up the entire execution time of PsC.AFLR, one can see that the time spent executing these operations increases as more cores are used and is executed simultaneously with PsC PDR and AFLR operations in about 90\% of the total runtime when run on 32 cores (showing that concurrency is indeed leveraged as more cores are utilized). When utilizing 64 cores, there is a decrease in load balancing and data dissemination. This is attributed to the fact that the number of subdomains remains fixed across all tested runs (execution with 1 core to execution with 64 cores), and there is not enough work for all the workers in this master/worker model used. Coupling this with the communication cost of using more than a single node on ODU's Turing cluster, increasing the number of cores utilized beyond 32 actually deteriorates performance. This could be alleviated in future work by increasing the PDR octree depth (to create more leaves) and by generating a denser initial mesh before decomposition.

\begin{figure}[h!]
\begin{center}
\includegraphics[width=0.45\textwidth]{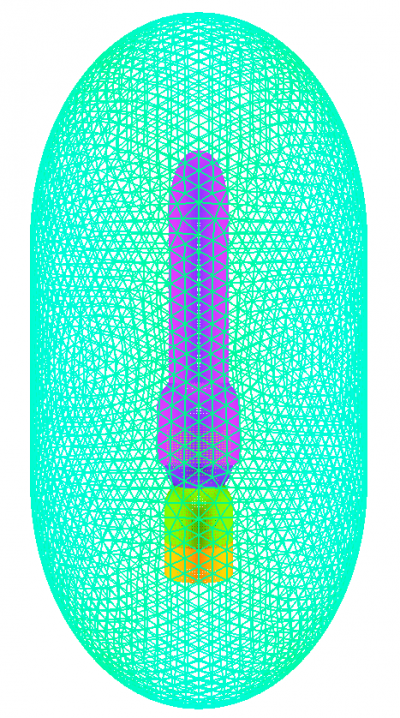}
\end{center}
\caption[Rocket Geometry]{Shown is a rocket geometry with 1,030,692 surface elements.}
\label{fig:rocket_pic}
\end{figure}

A rocket geometry, pictured in Figure \ref{fig:rocket_pic}, was also tested with both PsC.AFLR and serial AFLR. The rocket geometry originally contained transparent/embedded surfaces. These surfaces were specifically the plume, engine exhaust, and nozzle exhaust. Due to the limited capabilities of this implementation and the fact that PsC.AFLR currently only refines manifold, genus zero geometries, these surfaces were removed from the rocket geometry before testing. The rocket geometry contains 1,030,692 surface elements. The number of tetrahedra within the initial volume mesh generated by TetGen were 2,776,378 for PsC.AFLR. The octree used in PsC.AFLR is again at depth level 4 (containing 4,096 leaves, or subdomains). The final number of tetrahedra generated for the rocket geometry by PsC.AFLR is approximately 38 million and 146 million for serial AFLR. Serial AFLR’s refinement time is 16,646 seconds and its refinement speed is 8,785 elements/sec. Table \ref{table:rocket_perf_table} shows the performance achieved by PsC.AFLR, in addition to its refinement speed for each configuration of cores used, when refining the rocket geometry.

\begin{table}[h!]\scriptsize
\centering
\resizebox{\columnwidth}{!}{%
\begin{tabular}{ccccccccc}
 \hline
 \# of Cores&\multicolumn{2}{c}{1}&2&4&8&16&32&64\\
 \hline
 \# of generated elements & \textcolor{red}{146.2M} & 38M & 39.2M & 37.7M & 37.7M & 38M & 37.9M & 37.7M \\
 Runtime (sec) & \textcolor{red}{16.6K} & 1.6K & 1.2K & 1.4K & 1.1K & 1.1K & 1K & 1.3K \\
 Refine Speed (Elems/sec) & \textcolor{red}{8.7K} & 23.1K & 30.4K & 26.6K & 32.7K & 33.5K & 37.6K & 28.3K \\
 \hline
\end{tabular}%
}
\caption[PsC.AFLR Rocket Performance Results]{Approximate PsC.AFLR rocket performance results are shown. The serial AFLR data is denoted in red while the remainder is of PsC.AFLR. `K' means thousand and `M' means million.}
\label{table:rocket_perf_table}
\end{table}

\begin{figure}[h!]
\begin{center}
\includegraphics[width=0.8\textwidth]{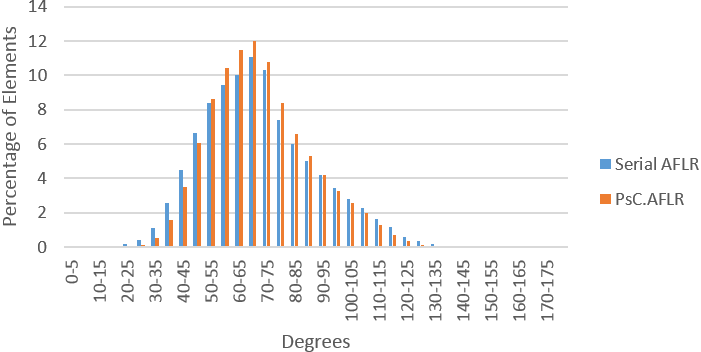}
\end{center}
\caption[Overall PsC.AFLR Dihedral Angle Statistics for Rocket]{The overall dihedral angle statistics of the rocket mesh generated by PsC.AFLR are shown.}
\label{fig:Rocket_Quality_Implementation_2_Overall}
\end{figure}

\begin{figure}[h!]
\begin{center}
\includegraphics[width=0.75\textwidth]{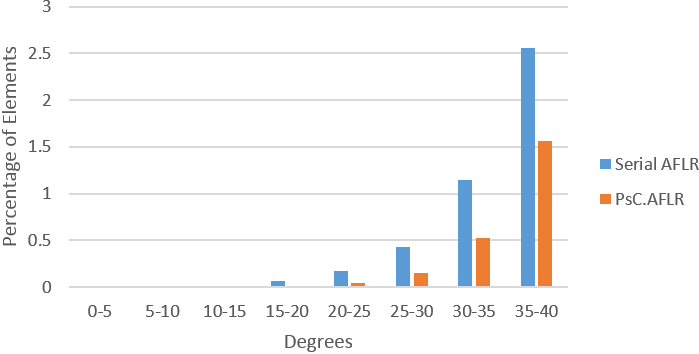}
\end{center}
\caption[Small PsC.AFLR Dihedral Angle Statistics for Rocket]{The small dihedral angle statistics of the rocket mesh generated by PsC.AFLR are shown.}
\label{fig:Rocket_Quality_Implementation_2_Small}
\end{figure}

\begin{figure}[h!]
\begin{center}
\includegraphics[width=0.75\textwidth]{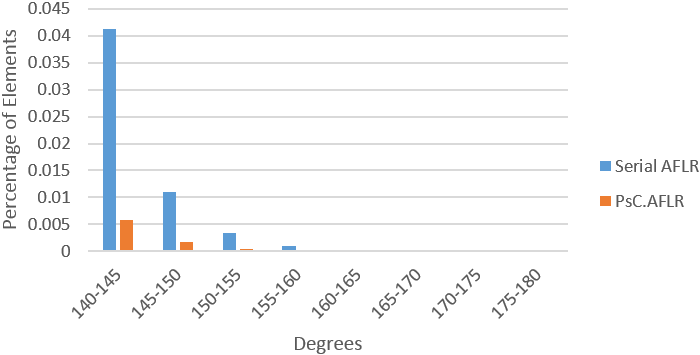}
\end{center}
\caption[Large PsC.AFLR Dihedral Angle Statistics for Rocket]{The large dihedral angle statistics of the rocket mesh generated by PsC.AFLR are shown.}
\label{fig:Rocket_Quality_Implementation_2_Large}
\end{figure}

Based on the results observed in Table \ref{table:rocket_perf_table}, PsC.AFLR outperforms serial AFLR (in both total runtime and refinement speed) and is capable of generating a larger number of elements per second than the serial code (even on one core, which will also be explained in the analysis in section \ref{analysis}). Although PsC.AFLR produces a final mesh with fewer elements than that produced by the serial software, its final mesh maintains satisfactory quality in comparison as shown in Figures \ref{fig:Rocket_Quality_Implementation_2_Overall}, \ref{fig:Rocket_Quality_Implementation_2_Small}, and \ref{fig:Rocket_Quality_Implementation_2_Large}. The minimum dihedral angle of an element in the final mesh is 2.27 degrees while the maximum is 164.97 degrees with PsC.AFLR (as opposed to serial AFLR having a minimum of 0.01 degrees and a maximum of 179.98 degrees).

\subsection{Analysis} \label{analysis}
Although it offers good end-user productivity, PsC.AFLR suffers in its capability (in both implementations) to generate meshes with the same level of density or quality as that of the serial AFLR software due to the constraints set by subdomain boundaries that are required to not only successfully execute AFLR but also determine its execution behavior given the nature of the advancing front method. PsC.AFLR and serial AFLR essentially produce two different final volume meshes when refining the same geometry. The final number of elements between the two applications is different for two reasons:
\begin{enumerate}
    \item PsC.AFLR extracts a surface triangulation for a subdomain based on the initial volume elements and these surface elements remain frozen throughout the refinement of that particular leaf. In the implementation that included super-subdomain local reconnection, the surface elements undergo local reconnection once the newly refined leaf data is merged with the elements within the surrounding level one neighbor leaves, but they do not undergo point insertion (which would create more elements). 
    \item Serial AFLR was run in its default state (no parameter adjustments). This is not satisfactory for PsC.AFLR when refining subdomains because it uses an initial mesh generated by TetGen (due to the aforementioned initial volume mesh generation challenges with AFLR) and therefore utilizes the point distribution of the subdomains' internal interface surfaces (from the coarse TetGen mesh) rather than the external dense surface of the geometry (which serial AFLR uses). This point distribution directly affects the number of elements created. As an attempted solution, AFLR’s point distribution function was abstracted to be called before refinement within PsC.AFLR so that a point distribution value based on the external surface is obtained. The point distribution for each subdomain is then explicitly set to this value. Henceforth, PsC.AFLR is able to generate a denser mesh of higher quality although it extracts coarse surfaces based on its initial TetGen-generated mesh. However, even with this adjustment, these constraints do not allow PsC.AFLR to generate a mesh as dense as that generated by the serial AFLR software and the resulting quality of subdomain volume elements is still constrained by the frozen internal interface elements.
\end{enumerate}

\begin{figure}[h!]
     \centering
     \begin{subfigure}[htb]{0.45\textwidth}
         \centering
         \includegraphics[width=\textwidth]{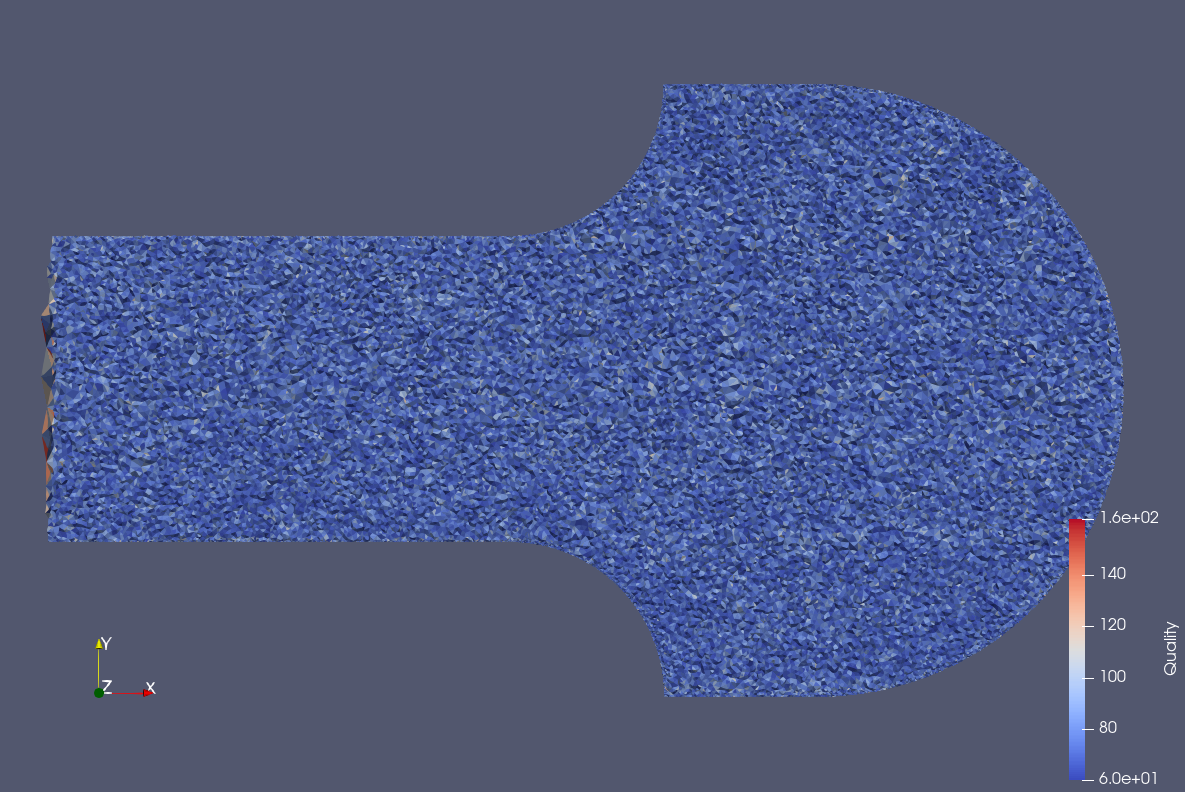}
         \caption{}
         \label{}
     \end{subfigure}
     \hfill
     \begin{subfigure}[htb]{0.45\textwidth}
         \centering
         \includegraphics[width=\textwidth]{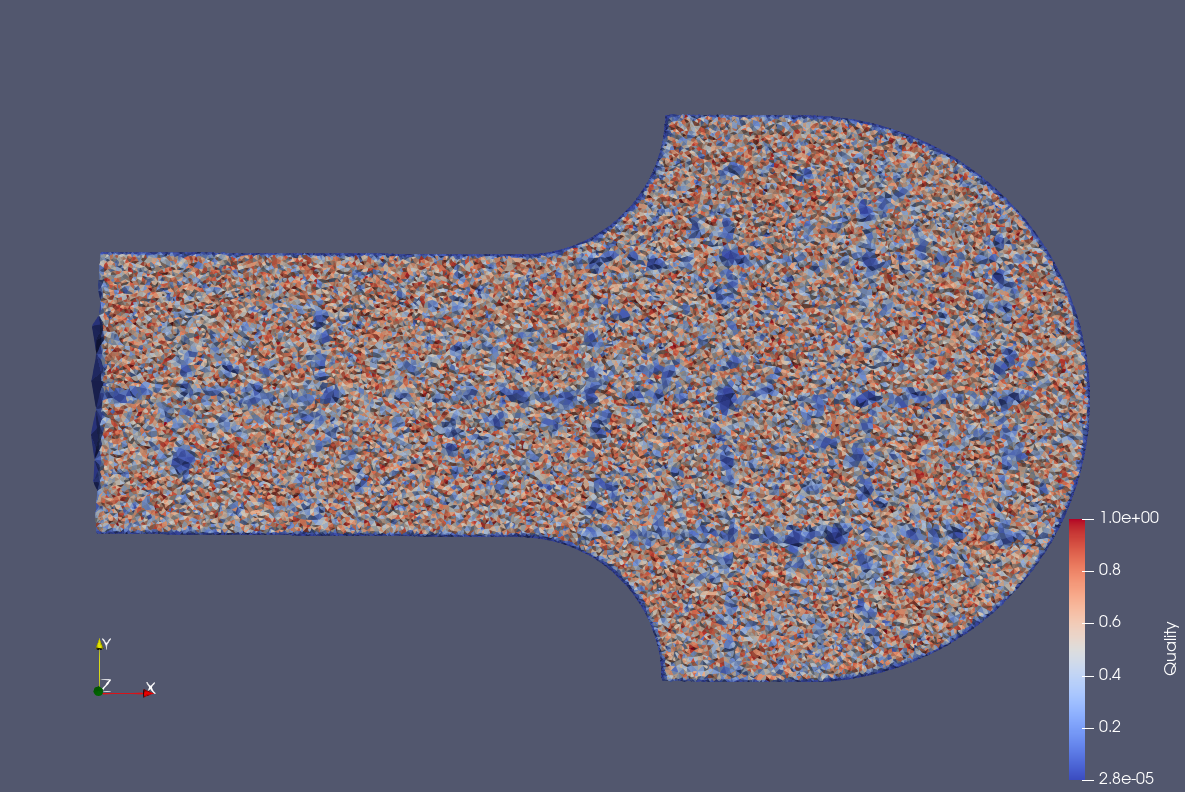}
         \caption{}
         \label{}
     \end{subfigure}
     \hfill
     \begin{subfigure}[htb]{0.45\textwidth}
         \centering
         \includegraphics[width=\textwidth]{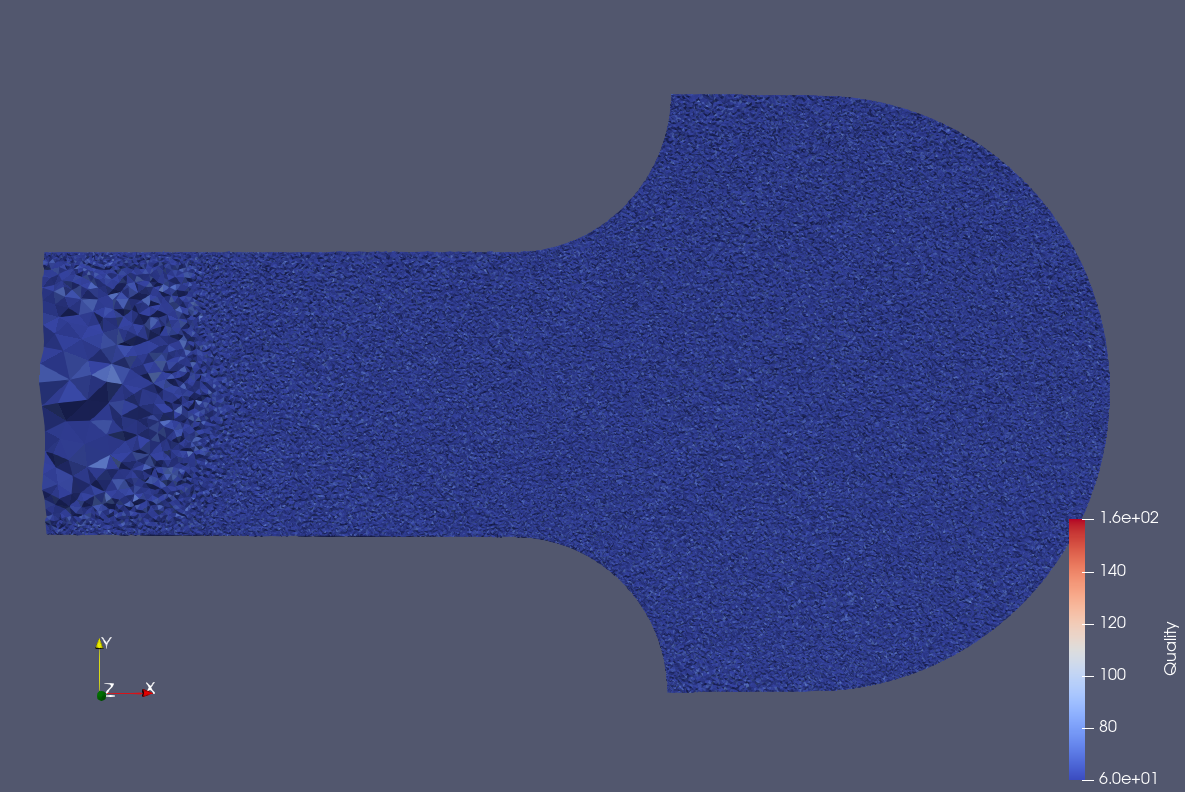}
         \caption{}
         \label{}
     \end{subfigure}
     \hfill
     \begin{subfigure}[htb]{0.45\textwidth}
         \centering
         \includegraphics[width=\textwidth]{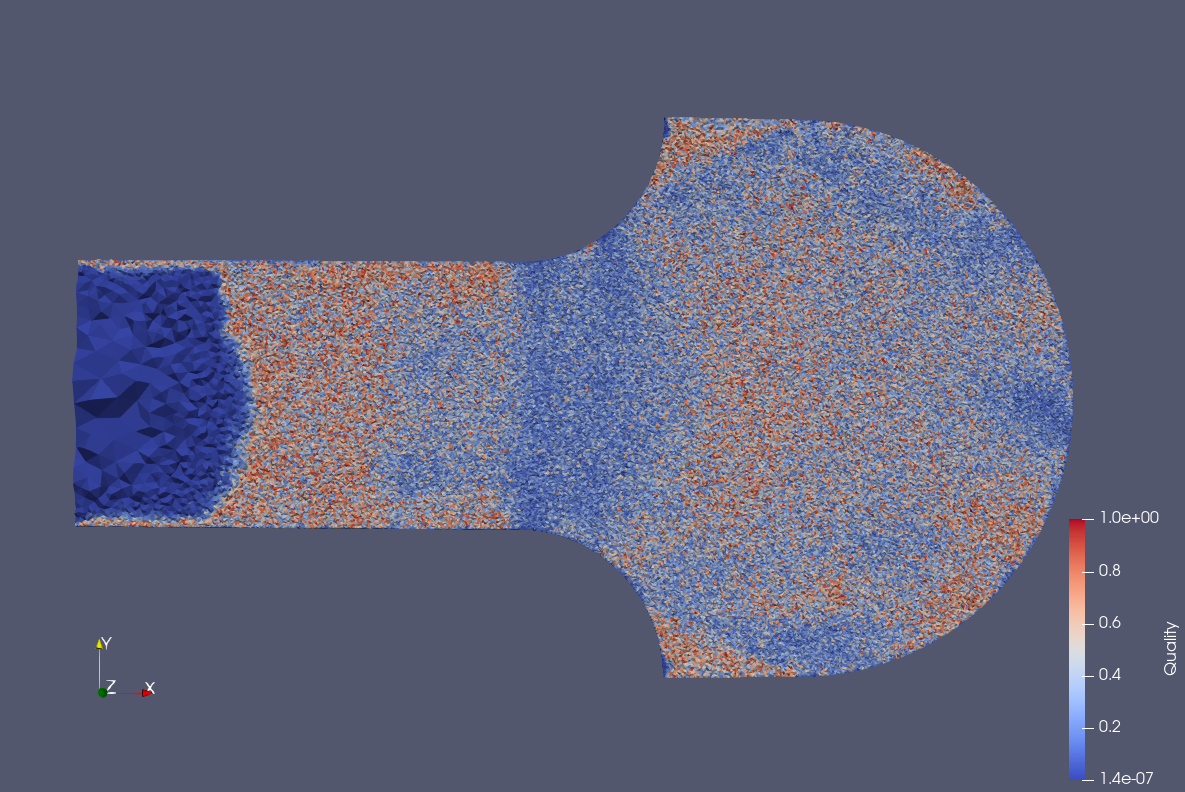}
         \caption{}
         \label{}
     \end{subfigure}
        \caption[PsC.AFLR and Serial AFLR Horn Bulb Slices Colored by Dihedral Angle and Shape \& Size Quality]{Dihedral Angle \& Shape and Size Quality Metrics Visualized. Slices of the horn bulb final meshes genereated by PsC.AFLR (a)(b) and serial AFLR (c)(d) are shown. (a) and (c) are colored by dihedral angle quality. (b) and (d) are colored by the shape and size quality metric defined in \cite{VerdictLibrary}.}
        \label{fig:Horn_Bulb_Quality_Slices}
\end{figure}

Further evidence of how the coarse boundary elements affect subdomain density can be seen in Figure \ref{fig:Horn_Bulb_Quality_Slices}, which shows dihedral angles plus shape and size quality metrics, as defined in \cite{VerdictLibrary}, between the volumes generated for the horn bulb geometry by PsC.AFLR (a and b) and serial AFLR (c and d). PsC.AFLR maintains good dihedral angle quality, as shown in Figure \ref{fig:Horn_Bulb_Quality_Slices}(a). Figure \ref{fig:Horn_Bulb_Quality_Slices}(b) shows the subdomain boundary elements, identified by their dark blue color, constraining the interior subdomain elements, thus preventing the subdomains from achieving the same density as the volume generated by serial AFLR in Figure \ref{fig:Horn_Bulb_Quality_Slices}(d). The minimum values for the shape and size metric are $2.8*10^{-5}$ in the mesh generated by PsC.AFLR and $1.4*10^{-7}$ in the mesh generated by serial AFLR. Serial AFLR is expected to have elements with a shape and size metric that is smaller due to its much larger density/volume of elements (8 times larger) and its better dihedral angle quality. One can observe that the density of the subdomains, within the volume generated by PsC.AFLR, is affected by the poor quality of the shape and relative size of the subdomain boundary elements. These internal interface elements constrain the interior elements (regardless of manually setting the target point distribution) and prevent the subdomains from becoming as dense as that seen in the volume generated by the serial AFLR method. This is also why PsC.AFLR seems to have better performance than serial AFLR even when executed on one core, because PsC.AFLR essentially generates a different, coarser volume. This issue stems from the fact that PsC.AFLR is dependent on the coarse surfaces extracted from the initial volume mesh. PsC.AFLR redefines the problem by having AFLR focus on subdomains (surrounded by these coarse surfaces) individually rather than on the entire domain. The elements within these subdomains are therefore refined differently (to a coarser level of density than that of the serial-output mesh in order to keep subdomain boundary elements frozen and to maintain their shape and relative size) than they would be if the domain was refined as a whole based on the original external surface.

\subsection{Lessons Learned}
Based on these results, the following abstractions can be made regarding the parallelization of a sequentially-designed black box mesh generation software like AFLR.
\begin{enumerate}
    \item One must be aware of how a black box software is designed to process input data when decomposing a problem into sub-problems. When a computation is dependent on a sequentially-designed black box, the processing of sub-problem inputs will not necessarily yield an equivalent solution to that of the original sequentially-processed problem input.
    \item Interface elements that do not already satisfy quality criteria can impose significant constraints on parallel mesh generation methods that rely upon an input surface geometry (for each subdomain) in order to execute advancing front point placement and local reconnection operations.
    \item In order to reach full stability, a parallel mesh generation method must be capable of satisfying target point distribution and quality criteria for elements within individual subdomains regardless of frozen interface elements (so that advancing front point placement and local reconnection can operate successfully in achieving the desired quality for those elements that are not immediately adjacent to interface elements).
    \item Unless the sequential black box software is redesigned, potential scalability may be hindered by pre- and post-refinement operations meant to prepare data for processing by the sequential black box software.
\end{enumerate}

\section{Future Work} \label{future_work}
Based on past work \cite{Chrisochoides18PDR} and results, an unconstrained parallel data refinement implementation is expected to be much faster and efficient than the pseudo-constrained PDR.AFLR. If one wished to further develop a parallel functionality-first approach with AFLR, several fundamental changes would need to be made to AFLR's source code in order to maintain the same level of functionality and achieve scalable performance when utilizing the concurrency offered by emerging large-scale HPC architectures. Most notably, these constraints would be alleviated if AFLR was capable of processing an input volume to a specified point distribution without requiring an input surface (such as the method in \cite{DrakopoulosCDT3D, EwCCDT3D}). Although failing the code re-use criterion (defined in section \ref{aflr_introduction}), implementing changes to AFLR while maintaining its functionality would also allow the parallel method to be more robust (as PsC.AFLR is currently only capable of processing manifold, genus zero geometries, as opposed to the robustness of the serial method). PsC.AFLR's robustness can be further improved by identifying octree leaves that contain disconnected volumes of a mesh (caused by hole(s) in the geometry, e.g., genus greater than zero) and these individual pieces can be refined independently of each other. With regards to geometries with transparent/embedded surfaces, an embedded surface must remain frozen during refinement. Because there is a volume on both sides of the surface, both sides can be considered as two separate subdomains and refined independently. The need to swap elements between subdomains (to ensure a smooth, simply-connected subdomain boundary) may be alleviated by utilizing a different decomposition technique altogether. A challenge with data decomposition methods is that they do not necessarily guarantee a decomposition with smooth, simply-connected subdomains. That being said, the mathematically proven optimization-based geometry partitioning algorithm in \cite{CHRISOCHOIDES199475} may reduce the number of disconnected partitions. One may also consider alternatives to the coarse, low quality subdomain boundary problem. The parallelized AFLR could take inspiration from the method in \cite{Lohner2014RecentAdvances}, which shifts layers of elements between subdomains to essentially allow interface elements to become interior elements. New interface elements (already refined) are frozen and these new subdomains are refined (permitting the refinement, including point insertion, of the original interface elements which are now interior). Such an algorithm involving AFLR should be carefully designed, however. This shifting of data was reported to hinder scalability not only in the functionality-first approach in \cite{Lohner2014RecentAdvances} but also in a scalability-first approach in \cite{KevinCDT3D}.

Given the complexity of such a state-of-the-art code (AFLR has over 100k lines of code developed over 30 years), the results of this 3-year effort encourages the development of a scalability-first approach rather than continued development with this constrained functionality-first approach. Furthermore, given the impact of subdomain boundary elements on the density and quality of individual subdomain volumes, a methodology should be utilized that successfully processes these interface elements and improves the quality of the output mesh without inducing significant overhead to satisfy constraints. Such a scalability-first method has been developed in chapter 3 of \cite{GarnerDissertation}, building upon a shared memory local reconnection-based method that was shown to have comparable output mesh quality to AFLR with optimal performance \cite{DrakopoulosCDT3D}. The scalability-first method in \cite{GarnerDissertation} showcased good performance by processing subdomain interface elements before any interior elements are adapted (taking inspiration from the approach in \cite{LeonidasDelaunayDecoupling2008}). If targeting large-scale architectures or beyond, reducing communication operations (to resolve data dependencies, such as shifting elements between subdomains) becomes fundamental to achieve good performance.

\section{Conclusion} \label{conclusion}
Due to the requirements and constraints set by AFLR, it is not possible to simply use it in a black box manner when attempting to parallelize the software. One must consider the overhead introduced simply from preparing a subdomain of data for refinement. Partition reassignment, boundary extraction, and data assignment all play a significant role in both the success of refinement (generating correct results, i.e., a conforming mesh with good quality) and in the runtime. Although they produce overhead on the runtime, these operations are essential when parallelizing AFLR if one wishes to minimize the modifications needed for AFLR (given its simply-connected input boundary requirement). PsC.AFLR has good end-user productivity given its refinement speed and is able to outperform serial AFLR by about 11 times when utilizing 16 CPU cores. The method maintains stability (in terms of quality), and AFLR was shown to satisfy the weak reproducibility criterion. Still, a pressing issue is that meshes generated by PsC.AFLR are much less dense (i.e., utilize a different point distribution) than those generated by serial AFLR and consequently may not have the same quality. The two programs essentially produce two different final volume meshes, regardless of attempting to run AFLR within PsC.AFLR using the same settings as serial AFLR (including when using identical point distribution values). AFLR's boundary requirement (extracted for each subdomain) constrain the capabilities of PsC.AFLR to generate dense meshes of high quality. These observations suggest that it is not practical to simply utilize a sequential code as a black box when developing a parallel mesh generation software with the functionality-first approach. There are unknown variables in the black box that can impede one from producing the desired results, suggesting that it may be pertinent for one to be more intimately familiar with the source code of the sequential mesh generator, which will enable them to make the necessary modifications to achieve these desired results (such as the methods in \cite{Lohner2014RecentAdvances} and \cite{MassivelyParallel2019}). On the other hand, this study encourages the development of a scalability-first approach (like the methods in \cite{DrakopoulosCDT3D} and \cite{GarnerDissertation}) that can produce results comparable to state-of-the-art methods while leveraging the concurrency offered by emerging HPC architectures to achieve good performance.


\section{Acknowledgements}
{
This research was sponsored in part by the NASA Transformational Tools and Technologies Project (NNX15AU39A) of the Transformative Aeronautics Concepts Program under the Aeronautics Research Mission Directorate, NSF grant no. CCF-1439079, the Richard T. Cheng Endowment, the Southern Regional Education Board (SREB) State Doctoral Scholar Fellowship, and the Virginia Space Grant Consortium Graduate Research Fellowship. The authors would like to thank Dr. Dana Hammond and Dr. Mike Park for their management as part of NASA's Transformational Tools and Technologies Project (NNX15AU39A) of the Transformative Aeronautics Concepts Program, and Dr. Park for imparting valuable knowledge regarding mesh quality assurance for CFD solvers. We would also like to thank Thomas Kennedy, Dr. Christos Tsolakis, and Dr. Polykarpos Thomadakis for their assistance in understanding how to utilize PDR and PREMA for this project.
}





\bibliographystyle{elsarticle-num} 
\bibliography{ltexpprt_references}

@inproceedings{ChernikovPDR2D2004,
author = {Chernikov, Andrey N. and Chrisochoides, Nikos P.},
title = {Practical and Efficient Point Insertion Scheduling Method for Parallel Guaranteed Quality Delaunay Refinement},
year = {2004},
isbn = {1581138393},
publisher = {Association for Computing Machinery},
address = {New York, NY, USA},
doi = {10.1145/1006209.1006217},
booktitle = {Proceedings of the 18th Annual International Conference on Supercomputing},
pages = {48–57},
numpages = {10},
location = {Malo, France},
series = {ICS '04}
}

@inproceedings{ChernikovPDR3D2008,
author = {Chernikov, Andrey N. and Chrisochoides, Nikos P.},
title = {Three-Dimensional Delaunay Refinement for Multi-Core Processors},
year = {2008},
isbn = {9781605581583},
publisher = {Association for Computing Machinery},
address = {New York, NY, USA},
doi = {10.1145/1375527.1375560},
booktitle = {Proceedings of the 22nd Annual International Conference on Supercomputing},
pages = {214–224},
numpages = {11},
location = {Island of Kos, Greece},
series = {ICS '08}
}

@article{TetGen2015,
author = {Si, Hang},
title = {TetGen, a Delaunay-Based Quality Tetrahedral Mesh Generator},
year = {2015},
issue_date = {January 2015},
publisher = {Association for Computing Machinery},
address = {New York, NY, USA},
volume = {41},
number = {2},
issn = {0098-3500},
doi = {10.1145/2629697},
journal = {ACM Transactions on Mathematical Software},
month = {02},
articleno = {11},
numpages = {36}
}

@article{Lohner2014RecentAdvances,
  title={Recent Advances in Parallel Advancing Front Grid Generation},
  author={Rainald L{\"o}hner},
  journal={Archives of Computational Methods in Engineering},
  year={2014},
  volume={21},
  issue={2},
  pages={127-140},
  doi={10.1007/s11831-014-9098-8}
}

@inproceedings{LohnerIMR2013,
author="L{\"o}hner, Rainald",
editor="Jiao, Xiangmin
and Weill, Jean-Christophe",
title="A 2nd Generation Parallel Advancing Front Grid Generator",
booktitle="Proceedings of the 21st International Meshing Roundtable",
year="2013",
publisher="Springer Berlin Heidelberg",
address="Berlin, Heidelberg",
pages="457-474",
isbn="978-3-642-33573-0"
}

@misc{ODUTuring,
author = {Old Dominion University},
title = {HPC Computational Clusters},
year = {2025},
howpublished = {\url{https://www.odu.edu/research-computing/compute}},
note = {\uppercase{O}nline; accessed 04/24/2025}
}

@misc{Fun3D,
author = {Beth Lee-Rausch},
publisher = {\uppercase{NASA}},
title = {\uppercase{FUN3D} Manual},
year = {2025},
howpublished = {\url{https://fun3d.larc.nasa.gov/}},
note = {\uppercase{O}nline; accessed 04/24/2025}
}

@article{SU2,
author = "Thomas D. Economon and Francisco Palacios and Sean R. Copeland and Trent W. Lukaczyk and Juan J. Alonso",
title = "\uppercase{SU}2: An Open-Source Suite for Multiphysics Simulation and Design",
journal = "AIAA Journal",
Volume = "54",
number = "3",
year = "2016",
month = "03",
doi = "10.2514/1.J053813"
}

@misc{VerdictLibrary,
author = {Stimpson, Clinton and Ernst, Corey and Knupp, P and Pébay, P and Thompson, Darby},
publisher = {Sandia National Laboratories},
year = {2007},
month = {04},
pages = {},
title = {The Verdict Library Reference Manual}
}

@inproceedings{CFD2030,
author = {Slotnick, Jeffrey P. and Khodadoust, Abdollah and Alonso, Juan and Darmofal, David and Gropp, William and Lurie, Elizabeth and Mavriplis, Dimitri J.},
title = {\uppercase{CFD} Vision 2030 Study: A Path to Revolutionary Computational Aerosciences},
year = {2014},
publisher = {NASA},
doi = {2060/20140003093},
location = {USA},
note = {\uppercase{N}ASA CR-2014-218178}
}

@article{MarcumAFLR,
author = "Marcum, David L. and Weatherill, Nigel P.",
title = "Unstructured Grid Generation Using Iterative Point Insertion and Local Reconnection",
journal = "AIAA Journal",
Volume = "33",
number = "9",
year = "1995",
month = "9",
page = "1619-1625",
doi = "10.2514/3.12701"
}

@inproceedings{Chrisochoides18PDR,
author = "Nikos Chrisochoides and Andrey Chernikov and Thomas Kennedy and Christos Tsolakis and Kevin Garner",
title = "Parallel Data Refinement Layer of a Telescopic Approach for Extreme-scale Parallel Mesh Generation for \uppercase{CFD} Applications",
year = "2018",
month = "6",
address = "Atlanta, Georgia",
booktitle = "2018 Aviation Technology, Integration, and Operations Conference",
note = "\uppercase{A}IAA 2018-2887",
doi = "10.2514/6.2018-2887"
}

@inproceedings{Chrisochoides2016TelescopicAF,
  title={Telescopic Approach for Extreme-Scale Parallel Mesh Generation for CFD Applications},
  author={Nikos Chrisochoides},
  year={2016},
  month={06},
  booktitle={46th AIAA Fluid Dynamics Conference},
  address={Washington D.C., USA},
  doi = {10.2514/6.2016-3181},
  note={\uppercase{A}IAA 2016-3181}
}

@article{DrakopoulosCDT3D,
author = "Drakopoulos, Fotis and Tsolakis, Christos and Chrisochoides, Nikos",
title = "Fine-Grained Speculative Topological Transformation Scheme for Local Reconnection Methods",
journal = "AIAA Journal",
Volume = "57",
number = "9",
year = "2019",
month = "9",
doi = "10.2514/1.J057657",
pages = "4007-4018"
}

@article{EwCCDT3D,
author = {Tsolakis, Christos and Chrisochoides, Nikos},
year = {2024},
month = {05},
pages = {3801–3827},
title = {Speculative Anisotropic Mesh Adaptation on Shared Memory for CFD Applications},
volume = {40},
issue = {},
journal = {Engineering with Computers},
doi = {10.1007/s00366-024-01994-0}
}

@article{ChernikovDistributedDelaunay2006,
author = { Chernikov, Andrey N. and  Chrisochoides, Nikos P.},
title = {Parallel Guaranteed Quality Delaunay Uniform Mesh Refinement},
journal = {SIAM Journal on Scientific Computing},
volume = {28},
number = {5},
pages = {1907-1926},
year = {2006},
doi = {10.1137/050625886}
}

@inproceedings{EPIC2012,
author = {Michal, Todd and Krakos, Joshua},
year = {2012},
month = {01},
pages = {},
title = {Anisotropic Mesh Adaptation Through Edge Primitive Operations},
isbn = {978-1-60086-936-5},
booktitle = {50th AIAA Aerospace Sciences Meeting Including the New Horizons Forum and Aerospace Exposition},
doi = {10.2514/6.2012-159},
note = {\uppercase{A}IAA 2012-0159}
}

@article{LOSEILLEFefloa,
title = {Parallel Generation of Large-size Adapted Meshes},
journal = {Procedia Engineering},
volume = {124},
pages = {57-69},
year = {2015},
note = {24th International Meshing Roundtable},
issn = {1877-7058},
doi = {10.1016/j.proeng.2015.10.122},
author = {Adrien Loseille and Victorien Menier and Frédéric Alauzet}
}

@inproceedings{ZagarisVGRID,
author = {Zagaris, George and Pirzadeh, Shahyar and Chrisochoides, Nikos},
year = {2013},
month = {06},
pages = {},
title = {A Framework for Parallel Unstructured Grid Generation for Practical Aerodynamic Simulations},
booktitle = {47th AIAA Aerospace Sciences Meeting including The New Horizons Forum and Aerospace Exposition},
address = {Orlando, Florida, USA},
doi = {10.2514/6.2009-980}
}

@article{NAVE2004191,
title = {Guaranteed-Quality Parallel Delaunay Refinement for Restricted Polyhedral Domains},
journal = {Computational Geometry},
volume = {28},
number = {2},
pages = {191-215},
year = {2004},
note = {\uppercase{S}pecial Issue on the 18th Annual Symposium on Computational Geometry - {S}o{CG}2002},
issn = {0925-7721},
doi = {10.1016/j.comgeo.2004.03.009},
author = {Démian Nave and Nikos Chrisochoides and L.Paul Chew}
}

@inproceedings{ParkRefine,
author = {Park, Michael and Darmofal, David},
year = {2008},
month = {01},
pages = {},
title = {Parallel Anisotropic Tetrahedral Adaption},
booktitle = {46th AIAA Aerospace Sciences Meeting and Exhibit},
address = {Reno, Nevada, USA},
doi = {10.2514/6.2008-917},
note = {\uppercase{A}IAA 2008-917}
}

@article{LeonidasDelaunayDecoupling2008,
author = {Linardakis, Leonidas and Chrisochoides, Nikos},
title = {Graded Delaunay Decoupling Method for Parallel Guaranteed Quality Planar Mesh Generation},
journal = {SIAM Journal on Scientific Computing},
volume = {30},
number = {4},
pages = {1875-1891},
year = {2008},
doi = {10.1137/060677276}
}

@ARTICLE{BarkerPREMA,  author={Barker, K. and Chernikov, A. and Chrisochoides, N. and Pingali, K.},  journal={IEEE Transactions on Parallel and Distributed Systems},   title={A Load Balancing Framework for Adaptive and Asynchronous Applications},   year={2004},  volume={15},  number={2},  pages={183-192},  doi={10.1109/TPDS.2004.1264800}}

@inproceedings{Thomadakis18PREMA,
author = "Polykarpos Thomadakis and Christos Tsolakis and Konstantinos Vogiatzis and Andriy Kot and Nikos Chrisochoides",
title = "Parallel Software Framework for Large-Scale Parallel Mesh Generation and Adaptation for \uppercase{CFD} Solvers",
booktitle = "AIAA Aviation Forum 2018",
year = "2018",
month = "6",
address = "Atlanta, Georgia",
note = "\uppercase{A}IAA 2018-2888",
doi = "10.2514/6.2018-2888"
}

@inproceedings{AmdahlsLaw,
author = {Amdahl, Gene M.},
title = {Validity of the Single Processor Approach to Achieving Large Scale Computing Capabilities},
year = {1967},
isbn = {9781450378956},
publisher = {Association for Computing Machinery},
address = {New York, NY, USA},
doi = {10.1145/1465482.1465560},
booktitle = {Proceedings of the April 18-20, 1967, Spring Joint Computer Conference},
pages = {483–485},
numpages = {3},
location = {Atlantic City, New Jersey},
series = {\uppercase{AFIPS} '67 (Spring)}
}

@article{APrioriEdgesIso2015,
author = {Soner, Seren and Ozturan, Can},
title = {Generating Multibillion Element Unstructured Meshes on Distributed Memory Parallel Machines},
journal = {Scientific Programming},
volume = {2015},
number = {1},
pages = {437480},
doi = {10.1155/2015/437480},
year = {2015}
}

@techreport{Pampa2017,
  TITLE = {{Fast Parallel Remeshing for Accurate Large-Eddy Simulations on Very Large Meshes}},
  AUTHOR = {Lachat, C{\'e}dric and Pellegrini, Fran{\c c}ois and Dobrzynski, C{\'e}cile and Staffelbach, Gabriel},
  TYPE = {Research Report},
  NUMBER = {RR-9133},
  PAGES = {13},
  INSTITUTION = {{Inria Bordeaux Sud-Ouest}},
  YEAR = {2017},
  MONTH = Dec,
  HAL_ID = {hal-01669775},
  HAL_VERSION = {v1},
}

@techreport{ParMMG2019,
  TITLE = {{Parallel Unstructured Mesh Adaptation Using Iterative Remeshing and Repartitioning}},
  AUTHOR = {Cirrottola, Luca and Froehly, Algiane},
  TYPE = {Research Report},
  NUMBER = {RR-9307},
  INSTITUTION = {{INRIA Bordeaux, {\'e}quipe CARDAMOM}},
  YEAR = {2019},
  MONTH = Nov,
  HAL_ID = {hal-02386837},
  HAL_VERSION = {v1},
}

@article{MassivelyParallel2019,
author = {Hugues Digonnet and Thierry Coupez and Patrice Laure and Luisa Silva},
title ={Massively Parallel Anisotropic Mesh Adaptation},
journal = {The International Journal of High Performance Computing Applications},
volume = {33},
number = {1},
pages = {3-24},
year = {2019},
doi = {10.1177/1094342017693906}
}

@misc{PDRPODMExperience,
title={Experience with Distributed Memory Delaunay-Based Image-to-Mesh Conversion Implementation}, 
author={Polykarpos Thomadakis and Nikos Chrisochoides},
year={2023},
eprint={2308.12525},
archivePrefix={arXiv},
primaryClass={cs.DC},
url={https://arxiv.org/abs/2308.12525}
}

@article{CHRISOCHOIDES199475,
title = {Mapping Algorithms and Software Environment for Data Parallel {PDE} Iterative Solvers},
journal = {Journal of Parallel and Distributed Computing},
volume = {21},
number = {1},
pages = {75-95},
year = {1994},
issn = {0743-7315},
doi = {10.1006/jpdc.1994.1043},
author = {N. Chrisochoides and E. Houstis and J. Rice}
}

@phdthesis{GarnerDissertation,
    title    = {On the Scalability of Anisotropic Mesh Adaptation on Distributed and Shared Memory Architectures for Numerical Approximations},
    school   = {Old Dominion University},
    author   = {Garner, Kevin},
    year     = {2025}, 
    doi      = {}
}

@inproceedings{KevinCDT3D,
 author = {Garner, Kevin and Tsolakis, Christos and Thomadakis, Polykarpos and Chrisochoides, Nikos},
 title = {Towards Distributed Semi-speculative Adaptive Anisotropic Parallel Mesh Generation},
 booktitle = {AIAA Aviation Forum and Ascend 2024},
 year = {2024},
 address = {Las Vegas, NV, USA},
 doi = {10.2514/6.2024-4505},
 note = {\uppercase{A}IAA 2024-4505}
}

@inproceedings{GarnerThesis,
    title    = {Parallelization of the Advancing Front Local Reconnection Mesh Generation Software Using a Pseudo-Constrained Parallel Data Refinement Method},
    school   = {Old Dominion University},
    author   = {Garner, Kevin},
    year     = {2020}, 
    doi      = {10.25777/appr-3169},
    note     = {\uppercase{M}aster's thesis, Old Dominion University}
}






\end{document}